%% file: main_aip.tex
\documentclass[%
 aip,
 amsmath,amssymb,
 reprint,%
]{revtex4-1}

\usepackage{graphicx}

\usepackage[colorlinks=true, linkcolor=blue, citecolor=blue, urlcolor=blue]{hyperref}
\usepackage{hyperref}
\usepackage{dcolumn}
\usepackage{bm}
\usepackage{color}

\usepackage[utf8]{inputenc}
\usepackage[T1]{fontenc}
\usepackage{mathptmx}
\usepackage{mathtools}
\usepackage{etoolbox}
\usepackage{comment}
\usepackage[english]{babel}

\makeatletter
\def\@email#1#2{%
 \endgroup
 \patchcmd{\titleblock@produce}
  {\frontmatter@RRAPformat}
  {\frontmatter@RRAPformat{\produce@RRAP{*#1\href{mailto:#2}{#2}}}\frontmatter@RRAPformat}
  {}{}
}%
\makeatother
\begin{document}

\preprint{AIP/123-QED}

\title{Automated optimization of force field parameters against ensemble-averaged measurements with Bayesian Inference of Conformational Populations}
\author{Robert M. Raddi}\noaffiliation
\author{Vincent A. Voelz\textsuperscript{*}}\noaffiliation
\affiliation{
Department of Chemistry, Temple University, Philadelphia, PA 19122, USA. \\
E-mails: rraddi@temple.edu and vvoelz@temple.edu \\
\textsuperscript{*}Corresponding author
}

\date{\today}

\begin{abstract}
Accurate force fields are essential for reliable molecular simulations. These models are refined against quantum mechanical calculations and experimental measurements, which are subject to random and systematic errors. Bayesian Inference of Conformational Populations (BICePs) is a reweighting algorithm that reconciles simulated ensembles with sparse or noisy observables by sampling the full posterior distribution of conformational populations and experimental uncertainty. In this method, a metric called the BICePs score is used to perform model selection, by calculating the free energy of "turning on" the conformational populations under experimental restraints.  This approach, when used with improved likelihood functions to deal with experimental outliers, can be used for force field validation (Raddi et al. 2025).  Here, we extend the BICePs approach to perform automated force field refinement while simultaneously sampling the full distribution of uncertainties, using a variational method to minimize the BICePs score. To demonstrate the utility of this method, we refine multiple interaction parameters for a 12-mer HP lattice model using ensemble-averaged distance measurements as restraints. To illustrate the resilience of BICePs in the presence of unknown random and systematic errors, we assess the performance of our algorithm through repeated optimizations and under various extents of experimental error. Our results suggest that variational optimization of the BICePs score is a promising direction for robust and automatic parameterization of molecular potentials.

\end{abstract}

\maketitle

\section{Introduction}
The accuracy of physical potentials used in molecular simulations continues to improve, especially as training data for fitting energy surfaces from quantum-mechanical calculations becomes more available.\cite{boothroyd2023development,smith2019approaching}  Of similar importance is developing models that can accurately predict ensemble-averaged experimental measurements. Parameterizing models for this purpose is a nontrivial task that requires continuous refinement of microscopic parameters through iterated sampling of the equilibrium distribution.\cite{wang2014building,poleto2022integration,smith2017ani,wang2018deepmd}


Such refinement often involves global minimization of the deviation between experimental measurements (e.g. NMR observables) and theoretical predictions from simulated ensembles (the ``forward model''), with several complications to consider. First, the forward model is itself an approximation containing some error. Second, ensemble-averaged experimental measurements can be sparse and/or noisy, susceptible to random and systematic errors, which are often unknown \textit{a priori}. Hence, force field refinement requires a mechanism to integrate these multiple sources of uncertainty to automatically discover the optimal parameter set across multiple systems.

Other challenges arise from practical considerations. The high dimensionality and interdependence of force field parameter space presents a significant challenge for refinement.  When parameter spaces are large, reproducibility may also be an issue; ideally, identical experimental data should produce parameterizations that should converge towards the same optimal values.  For these reasons, automated optimization procedures with sufficiently smooth and differentiable objective functions are preferred.




Numerous methods\cite{kofinger2021empirical,wang2014building,ge2018model,madin2022bayesian} have been developed to address most of these challenges. However, many algorithms do not include any treatment of uncertainty in the training data, and others lack gradients for automatic refinement.  To address these challenges, Bayesian inference methods have been developed.\cite{kofinger2021empirical,madin2022bayesian,ge2018model} These methods estimate a Bayesian posterior distribution of conformational populations by treating molecular simulation predictions as prior information, weighted by a likelihood function constructed from the experimental measurements and their uncertainties. The Bayesian posterior can then be used to optimize the prior.

In this work, we extend the Bayesian Inference of Conformational Populations (BICePs)  algorithm\cite{voelz2021reconciling,raddi2025model} to perform automated force field refinement.
BICePs is a maximum-entropy (MaxEnt) reweighting algorithm that refines structural ensembles against sparse and/or noisy experimental observables. Recent developments of BICePs enabled automatic parameterization of forward models\cite{Raddi2024FMO,NRV2024}, objective ranking of force fields\cite{raddi2025model}, and reweighting folding landscapes for non-natural and/or cyclic peptides\cite{NRV2024}. Earlier versions of BICePs used a maximum-parsimony approach to reconcile simulations with experimental data in peptide and protein systems\cite{Voelz:2014fga,wan2020reconciling,hurley2021,raddi2023biceps}.
A key advantage of BICePs is that it does not require uncertainty estimates for the forward model; instead, it infers the posterior distribution of these parameters directly from the data through MCMC sampling. BICePs also computes a free energy-like quantity called the BICePs score that can be used for model selection.\cite{ge2018model,raddi2025model}

Recently, we have equipped BICePs with a replica-averaging forward model, making it a MaxEnt reweighting method. Among existing MaxEnt methods, BICePs is unique in that no adjustable regularization parameters are required to balance experimental information with the prior.\cite{raddi2025model}  With this new approach, the BICePs score becomes a powerful objective function to select and parameterize optimal models.   Here, we show that the BICePs score, which reflects the total evidence for a model, can be used for variational optimization of model parameters.  The BICePs score contains a form of inherent regularization, and has specialized likelihood functions that allow for the automatic detection and down-weighting of data points subject to systematic error.\cite{raddi2025model}

To efficiently optimize complex parameter spaces, we derive the first and second derivatives of the BICePs score. We then show how to perform automatic force field optimization against ensemble-averaged observables for several test systems, including a protein lattice model with adjustable bead interaction strengths, and a von Mises-distributed polymer model. We also show that this approach works with neural network potentials, where parameters can be optimized through automatically calculated gradients.

\section{Theory}
BICePs uses a Bayesian statistical framework to treat the extent of uncertainty in experimental observables, $\sigma$, as nuisance parameters. Previous versions of BICePs sampled conformational states $X$ and uncertainty parameter(s) $\sigma$ from the Bayesian posterior, which takes the form
\begin{equation}
\overbrace{p(X,\sigma | D)}^{\text{posterior}} \propto  \overbrace{p(D | X,\sigma)}^{\text{likelihood}} \overbrace{p(X) p(\sigma)}^{\text{priors}} .
\end{equation}
Here, the prior $p(X)$ comes from a theoretical model of conformational state populations (typically from a molecular simulation), $p(D | X,\sigma)$ is a likelihood function quantifying how well a forward model prediction $f(X)$ agrees with the experimental data $D$, and $p(\sigma) \sim \sigma^{-1}$ is a non-informative Jeffreys prior.

When BICePs is equipped with a replica-averaged forward model, it becomes a MaxEnt reweighting method in the limit of large numbers of replicas \cite{pitera2012use,cavalli2013molecular,cesari2018using,roux2013statistical,hummer2015bayesian}.  The posterior takes the general form
\begin{equation}
  \begin{split}
    p(&\mathbf{X},\bm{\sigma}| D) \propto \\
    &\prod_{r=1}^{N_r} \Bigl\{ p(X_r) \prod_{j=1}^{N_j} \frac{1}{\sqrt{2\pi \sigma_j^2}} \exp \Big[ -\frac{(d_j - f_j(\mathbf{X}))^2}{2\sigma_j^2} \Big] p(\sigma_j) \Bigr\} \label{replica_posterior}
  \end{split}
\end{equation}
where $\mathbf{X}$ is a set of $N_r$ conformation replicas, $d_j$ is an observable in the set of $N_j$ ensemble-averaged experimental measurements, and $f_j(\mathbf{X}) = \frac{1}{N_r} \sum_r^{N_r} f_j(X_r)$ is the replica-averaged forward model prediction of observable $j$.  The $\sigma_j$ values are nuisance parameters that capture uncertainty in the measurements as well as the replica-averaged forward model. In \eqref{replica_posterior}, a Gaussian likelihood is used, but more sophisticated models of uncertainty can be used to capture outliers and systematic error with fewer parameters, as discussed below. Markov chain Monte Carlo (MCMC) is used to sample the posterior.

The replica-averaged forward model $f(\mathbf{X})$ requires a more sophisticated treatment of uncertainty, since it is approximating the ensemble average as a replica average.\cite{bonomi2016metainference}. Therefore,  $\sigma_j$ becomes a combination of both the Bayesian error $\sigma^{B}_{j}$ and the standard error of the mean $\sigma^{\text{SEM}}_{j}$, $\sigma_{j} = \sqrt{(\sigma^{\text{B}}_{j})^{2} + (\sigma^{\text{SEM}}_{j})^{2}}$. The finite sampling error, $\sigma^{\text{SEM}}$ is estimated by taking a windowed average over our finite sample $f(\textbf{X})$ as $\sigma^{SEM}_{j} = \sqrt{\frac{1}{N}\sum\nolimits^{N}_{r} (f_{j}(X_{r}) - \langle f_{j}(\mathbf{X}) \rangle )^{2}}$, which decreases as the square root of the number of replicas.

\subsection*{Accounting for systematic error and outliers}
BICePs is equipped with special likelihoods that are very robust in the presence of systematic error. The \textit{Student's} likelihood model is a data error model that marginalizes the uncertainty parameters for individual observables, assuming that the level of noise is mostly uniform, except for a few erratic measurements. This limits the number of uncertainty parameters that need to be sampled, while still capturing outliers.

The derivation of the Student's model proceeds as follows: Consider a model where uncertainties $\sigma_j$ for particular observables $j$ are distributed about some typical uncertainty $\sigma_{0}$ according to a conditional probability $p(\sigma_j | \sigma_{0})$. We derive a posterior with a single uncertainty parameter $\sigma_{0}$ by marginalizing over all $\sigma_j$. For a single replica (for simplicity), the posterior is given by
\begin{equation}
p(X_{r},\sigma_{0}|D) \propto p(X_{r}) \prod^{N_{j}}_{j=1} \int\limits_{\sigma^{\text{SEM}}}^{\infty} p(d_{j}|\mathbf{X}, \sigma_{j}) p(\sigma_{j} | \sigma_{0}) d\sigma_{j}
\label{eq:posterior_general_outliers}
\end{equation}
where  $\sigma_{0} = \sqrt{(\sigma^{\text{B}})^{2} + (\sigma^{\text{SEM}})^{2}}$.

Modeling the prior on uncertainties $p(\sigma_{j}|\sigma_0)$ with a distribution with long tails is very useful because its long tail makes it able to tolerate outliers\cite{sivia2006data,bonomi2016metainference}.  In most cases, however, it is unclear \textit{a priori} what distribution is best for modeling the input data. To improve the situation, we introduce a model with an additional nuisance parameter $\beta$, that is able to tune the extent of the distribution's tail:
\begin{equation}
  p(\sigma_{j}| \sigma_{0}, \beta) = \frac{\Gamma((\beta+1)/2)}{\Gamma(\beta/2)} \frac{2\beta^{\beta} {\sigma_{0}}^{2\beta-1}}{\sqrt{\beta} {\sigma_{j}}^{2\beta}} \exp \left(-\frac{\beta{\sigma_{0}}^{2}}{\sigma_{j}^{2}}\right).
  \label{eq:prior_students}
\end{equation}
where $\sigma_{0}$ is defined as above, and $1 \leq \beta < \infty$.  When this distribution is inserted into the posterior, and marginalized over all $\sigma_j$, the result is

\begin{equation}
  \begin{split}
    &p(X_{r}, \sigma_{0}, \beta | D) \propto p(X_{r}) \prod^{N_{j}}_{j=1} \int\limits_{\sigma^{\text{SEM}}}^{\infty} \frac{1}{\sqrt{2\pi}\sigma_{j}}  \exp\left( - \frac{(d_{j} - f_{j}(\mathbf{X}))^{2}}{2\sigma_{j}^{2}} \right)\\
    &\quad \quad \quad \times \frac{\Gamma((\beta+1)/2)}{\Gamma(\beta/2)} \frac{2\beta^{\beta} {\sigma_{0}}^{2\beta-1}}{\sqrt{\beta} {\sigma_{j}}^{2\beta}} \exp \left(-\frac{\beta{\sigma_{0}}^{2}}{\sigma_{j}^{2}}\right) d\sigma_{j} \\
    &=  p(X_{r}) \prod^{N_{j}}_{j=1}  \frac{\Gamma((\beta+1)/2)}{\Gamma(\beta/2)} \frac{1}{\sqrt{2\pi\beta} \sigma_{0}} \\
    &\quad\times\left[1 + \frac{(d_{j}-f_{j}(X))^{2}}{2 \beta \sigma_{0}^{2}}\right]^{-\beta} \gamma\left(\beta, \frac{(d_{j}-f_{j}(X))^{2} + 2 \beta \sigma_{0}^{2}}{2 \left(\sigma^{\text{SEM}}\right)^{2}}\right),
\end{split}
  \label{eq:posterior_students}
\end{equation}
where  $\gamma$ is the lower incomplete gamma function.  We call this the Student's model because it is a variation of Student's $t$-distribution that can be interpolated between functional forms. When $\beta=1$, the model is equivalent to Metainference's Outliers model\cite{bonomi2016metainference}.  In the limit of $\beta \rightarrow \infty$, the likelihood becomes Gaussian. When considering the full posterior, this extra nuisance parameter is given a non-informative Jeffreys prior,  $p(\beta) \sim \beta^{-1}$.
For a detailed solution to the marginalization integral in equation \ref{eq:posterior_students}, see Appendix \ref{sec:marginal_integral}.
For more details regarding the Student's model, including plots of the Student's prior and the likelihood probability distribution function for a wide range of $\beta$ values, see Figures \ref{fig:Students_restraint_marginal_likelihood}-\ref{fig:Students_prior_example}.

\subsection*{The BICePs score: a tool for quantitative model selection and refinement}

\begin{figure}
\centering
  \includegraphics[width=0.95\linewidth]{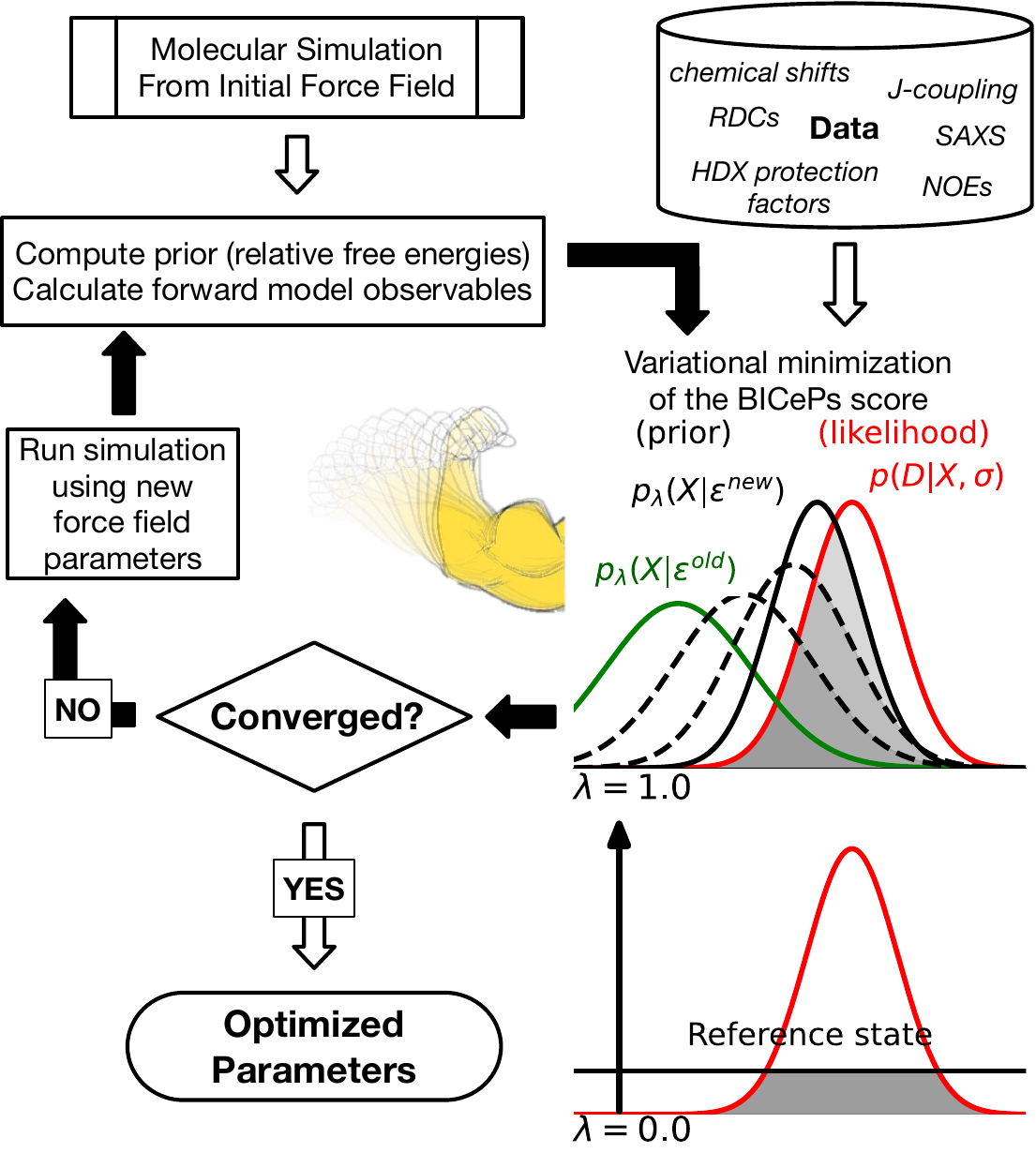}
  \caption{\small Force field optimization by  variationally minimizing the BICePs score. Given  ensemble-averaged experimental observables $D$, and a simulated ensemble $p_{\lambda}(X|\epsilon)$ generated from molecular simulation using initial force field parameters $\epsilon$ , an automated procedure is used to find the optimal parameters $\epsilon^{*}$ that best match experiment.  The procedure is guided by the first and second derivatives of the BICePs score at each iteration to propose new values of $\epsilon$.  This cycle is iterated until convergence is reached.
  }
  \label{fig:flowchart}
\end{figure}

The BICePs score is a free energy-like quantity that rigorously characterizes model quality. For a model of prior populations $p(X| \epsilon)$ parameterized by $\epsilon$, with corresponding prior energy $E(X| \epsilon)=-\ln p(X| \epsilon)$, the BICePs score $f(\epsilon)$ is calculated as the negative logarithm of a Bayes factor,
\begin{equation}
  f(\epsilon) = - \ln \left(Z(\epsilon) \big/ Z_{0}\right).
\end{equation}
$Z(\epsilon)$ is the total evidence of the specified model, marginalized over all conformational replicas and uncertainty. By defining the BICePs energy function as $u = -\ln p(\mathbf{X},\boldsymbol{\sigma}|D, \epsilon)$, it takes the form
\begin{equation}
  Z(\epsilon) = \int  \int   \exp\left(-u(\mathbf{X}, \mathbf{\sigma} \mid D, \epsilon)\right) d \mathbf{X} d \mathbf{\sigma},
\end{equation}
$Z_{0}$ is defined similarly, but for a uniform prior that does not depend on $\epsilon$.  $Z_{0}$ thus serves as a well-defined reference state. By constructing a series of intermediates that scale the prior $p_{\lambda}(X|\epsilon) \sim [p(X|\epsilon)]^{\lambda}$ as $\lambda=0 \rightarrow 1$, the BICePs score is calculated as the change in free energy for ``turning on'' the prior (Figure \ref{fig:flowchart}).

\subsection*{Variational minimization of the BICePs score to find optimal model parameters}
\label{sec:variational}

We assume there exists some posterior distribution $p^{*}(\mathbf{X}, \mathbf{\sigma} | D, \epsilon^{*})$ with the maximal evidence, derived from optimal parameters $\epsilon^{*}$. In this case, the free energy $f(\epsilon^{*})=f^{*}$ is a global minimum, and any other set of parameters $\epsilon$ has $f(\epsilon) - f^{*} \geq 0$.

Our objective is to find $\epsilon^{*}$ starting with trial parameters $\epsilon$ and corresponding distribution $p(\mathbf{X}, \mathbf{\sigma} | D, \epsilon)$.  To do this, we iteratively perform BICePs calculations $\lambda=0 \rightarrow 1$ to estimate $f(\epsilon)$, then update the parameters to further minimize this quantity, until the stopping criteria is reached.  Given a differentiable prior, $f(\epsilon)$ is also differentiable, enabling application of any number of gradient-based convex optimization schemes, with a stopping criteria of $\frac{\partial f}{\partial \epsilon} \big|_{\epsilon=\epsilon^{*}} = 0$, within some tolerance.

The derivative of the BICePs score with respect to parameters $\epsilon$ reduces to the ensemble-averaged value of $\partial u / \partial \epsilon_i$,
\begin{equation}
\begin{split}
  \frac{\partial f}{\partial \epsilon_{i}} &= \iint \frac{1}{Z(\epsilon)}  \left[ \frac{\partial u}{\partial \epsilon_{i}} \right] \exp \left( - u \right)  d \mathbf{X} d \mathbf{\sigma} = \bigg\langle \frac{\partial u}{\partial \epsilon_{i}} \bigg\rangle,
\end{split}
  \label{eq:df}
\end{equation}
where
\begin{equation}
  \begin{split}
    \frac{\partial u}{\partial \epsilon_{i}} = \sum^{N}_{r=1} \biggl\{ \frac{\partial E(X_{r} | \epsilon)}{\partial \epsilon_{i}} - \bigg\langle \frac{\partial E(\mathbf{X} | \epsilon)}{\partial \epsilon_{i}} \bigg\rangle \biggr\}
  \end{split}
  \label{eq:du}
\end{equation}
The reference ensemble $Z_0$ does not affect the gradient of $f$, since a uniform prior is independent of $\epsilon$, and $\langle \partial u_{0} / \partial \epsilon \rangle$ is zero (see section \ref{sec:SI_theory} for more details).

The second partial derivatives of the BICePs score with respect to parameters $\epsilon_i$ and $\epsilon_j$ are:
\begin{equation}
  \begin{aligned}
    \frac{\partial^{2} f(\epsilon)}{\partial \epsilon_{i} \partial \epsilon_{j}} &= \bigg\langle \frac{\partial^{2} u }{\partial \epsilon_{i}\partial \epsilon_{j}} \bigg\rangle
    - \bigg\langle  \frac{\partial u }{\partial \epsilon_{i}} \cdot \frac{\partial u }{\partial \epsilon_{j}} \bigg\rangle
    +\bigg\langle \frac{\partial u }{\partial \epsilon_{i}} \bigg\rangle \bigg\langle\frac{\partial u }{\partial \epsilon_{j}} \bigg\rangle \\
    &= \left\langle \frac{\partial^{2} u}{\partial \epsilon_{i}\partial \epsilon_{j}} \right\rangle - \operatorname{Cov} \!\left( \frac{\partial u}{\partial \epsilon_i}, \frac{\partial u}{\partial \epsilon_j}\right) .
  \end{aligned}
\end{equation}
The first term on the right is the ensemble-averaged second derivative of the energy function $u$ with respect parameters $\epsilon_{i}$ and $\epsilon_{j}$:
\begin{equation}
  \begin{split}
    \frac{\partial^{2} u }{\partial \epsilon_{i}\partial \epsilon_{j}} = \sum^{N}_{r=1} \biggl\{& \frac{\partial^{2} E(X_{r} | \epsilon)}{\partial \epsilon_{i}\partial \epsilon_{j}}
        - \bigg\langle \frac{\partial^{2} E}{\partial \epsilon_{i}\partial \epsilon_{j}} \bigg\rangle \\
        &+ \operatorname{Cov} \!\left( \frac{\partial E}{\partial \epsilon_i}, \frac{\partial E}{\partial \epsilon_j} \right) \biggr\}.
  \end{split}
\end{equation}

When $i$ = $j$, the second partial derivative of the BICePs score reduces to the difference between the ensemble-averaged second derivative of the energy $u$ and the variance of its first partial derivative:
\begin{equation}
  \begin{split}
    \frac{\partial^{2} f}{\partial \epsilon^{2}} &= \bigg\langle \frac{\partial^{2} u }{\partial \epsilon^{2}} \bigg\rangle
    - \left( \bigg\langle \left( \frac{\partial u }{\partial \epsilon}\right)^{2} \bigg\rangle
    -\bigg\langle \frac{\partial u }{\partial \epsilon} \bigg\rangle^{2} \right)
  \end{split}
  \label{eq:d2f}
\end{equation}

In practice, all of these quantities are estimated
as ensemble-averaged observables along with the free energy estimation of $f(\epsilon)$.  This is done using the MBAR free energy estimator, where the input consists of MCMC sampling at several intermediates $\lambda= 0\rightarrow 1$, to enable accurate estimates of all quantities (Figure \ref{fig:flowchart}).

\section{Methods}
\subsection*{A toy protein lattice model}
\begin{figure}
\centering
  \includegraphics[width=\linewidth]{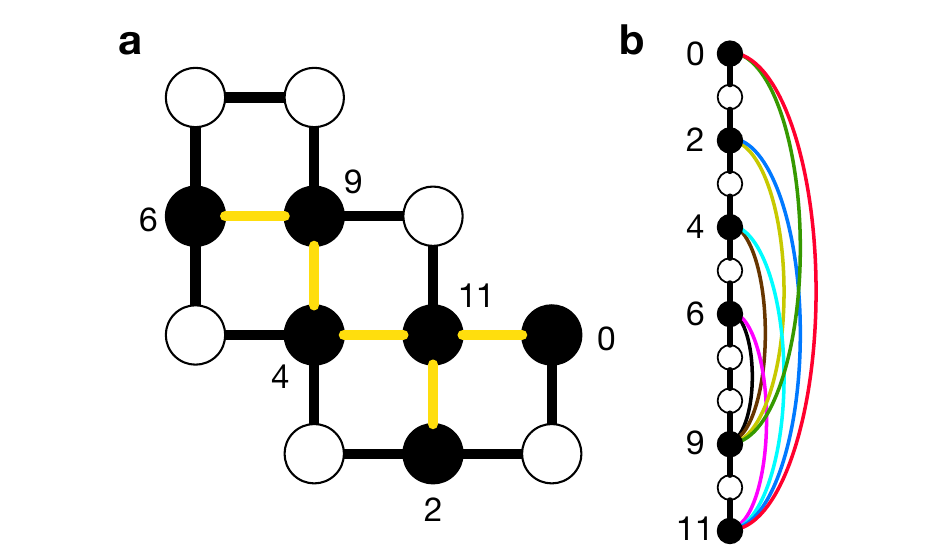}
  \caption{\small (a) The folded state of the 12-mer HP lattice model protein with sequence HPHPHPHPPHPH. Highlighted in yellow are the favorable non-bonded contacts between the hydrophobic residues (black beads). (b) Extended diagram of the eight distance measurements used in this work.}
  \label{fig:lattice_model}
\end{figure}

The HP lattice model of Dill and Chan\cite{dill1995principles} is a simplified protein folding model that represents a protein sequence as a chain of beads on a 2-D square lattice. In this model, each amino acid residue is classified as either a hydrophobic (H) or polar (P) bead. Each microscopic chain conformation $X$ has an energy $V(X)$ proportional to the number of H-H contacts $n(X)$,
\begin{equation}
V(X) = -\epsilon \cdot n(X),
\end{equation}
A ``contact'' is defined as a non-sequential pair of hydrophobic residues at adjacent lattice sites. As a test system, we consider a 12-mer protein sequence HPHPHPHPPHPH (Figure \ref{fig:lattice_model}a). When $\epsilon$ is sufficiently large, there is a single lowest-energy microstate that corresponds to the global free energy minimum; such sequences are deemed \textit{foldable}.  For the 12-mer HPHPHPHPPHPH sequence, there are 15037 unique microstate conformations that can be binned into 72 macrostates, each with a unique set of contacts $\mathcal{C}$.

To explore a system with multiple parameters, we modify the HP model to include multiple interactions.  Each bead with sequence position $i$ is assigned a unique interaction strength, $\epsilon_i$, and a geometric-average combination rule is used to compute the energy $V$ of each microstate:
\begin{equation}
  V = \sum_{{i,j} \in\mathcal{C}} -(\epsilon_i \epsilon_j)^{1/2},
\end{equation}
where the sum is taken over all pairs of residues $(i, j)$ that are in contact.

The first derivative of the contact energy with respect to $\epsilon_i$ is given by
\begin{equation}
\frac{\partial V}{\partial \epsilon_i} = \sum_{{i,j} \in\mathcal{C}} -\frac{1}{2} (\epsilon_i \epsilon_j)^{-1/2} \epsilon_j,
\end{equation}
and the second derivative of the contact energy with respect to $\epsilon_i$ and $\epsilon_j$ is given by:
\begin{equation}
\frac{\partial^2 V}{\partial \epsilon_i\partial \epsilon_j} =
\begin{dcases}
-\frac{1}{4}(\epsilon_i\epsilon_j)^{-1/2} & i\neq j,\ (i,j)\in\mathcal{C} \\
\sum_{(i,k)\in\mathcal{C}} \frac{1}{4}\frac{\epsilon_k}{(\epsilon_i\epsilon_k)^{3/2}} & i=j
\end{dcases}
\end{equation}
We use the 72 macrostates with unique contacts as a set of prior populations for use with BICePs.  For each macrostate $k$, the  populations $p_k$ are computed as
\begin{equation}
p_k = \frac{g_k \exp(-V_k)}{\sum_k g_k \exp(-V_k)},
\end{equation}
where $V_k$ is the reduced energy (in units $k_BT$) of each microstate in macrostate $k$, and  $g_k$ is the multiplicity (the number of microstates belonging to macrostate $k$).

The ensemble-averaged distance observables for each macrostate (and their variance over all microstates belonging to it) are computed as follows. Let $(d_{ij})_k$ represent the average distance between residues $i$ and $j$ in the $k^{th}$ macrostate. Then, the ensemble-averaged distance observables are given by:
\begin{equation}
\langle d_{ij} \rangle = \sum_k p_k (d_{ij})_k,
\end{equation}
and the variance of the distance observables is given by:
\begin{equation}
\sigma^2(\langle d_{ij}\rangle) = \sum_k p_k^2 \sigma^2((d_{ij})_k),
\end{equation}
where $\sigma^2((d_{ij})_k)$ is the variance of the average distance between residues $i$ and $j$ in the $k^{th}$ macrostate.

For ease of interpretation, we assume that the distance units of the square lattice are L.U..

\section{Methods}
\subsection*{Algorithms and settings for SciPy optimization}

With the ability to compute first and second derivatives of the BICePs score, we next explored the performance of automated optimization methods L-BFGS-B\cite{byrd1995limited}  and Trust-NCG\cite{wright2006numerical} on the single-$\epsilon$ system.  L-BFGS-B is a first-order approach that uses information from gradients, while Trust-NCG is a second-order approach that additionally makes use of information from the Hessian. All default optimization parameters from SciPy\cite{2020SciPy-NMeth} were used except for the value of \texttt{ftol=1e-8} (default: \texttt{ftol=1e-9}), and the bounds for each parameter to be optimized was set to be in the range [0, 10]. With only a single parameter, automatic refinement using both optimization methods quickly finds the minimum in a single iteration (Figure \ref{fig:bootstrapped_1-D_optimization}). However, use of the Hessian (second-order) enables faster convergence, and tends to be slightly more robust for this particular example.

\subsection*{BICePs score landscape visualization}

To visualize the two-parameter BICePs score landscape $f(\epsilon_2, \epsilon_4)$, we constructed a smooth 2-D landscape by averaging five repeated evaluations over a (6$\times$6) set of grid points, and trained a Gaussian process regression on the grid values using a radial basis function (RBF) kernel with an additive white noise to account for observational uncertainty.  The characteristic length scale was bounded within $[ 0.1, 10.0]$ and the signal variance was set to 1.0 and the noise variance was set to $10^{-5}$.

\subsection*{Estimation of parameter uncertainties with the inverse Hessian}

The reported uncertainties $\sigma_{\epsilon}$ in the optimized $(\epsilon_2, \epsilon_4)$ values come from an estimation of the covariance through inversion of the Hessian (the matrix of second partial derivatives of the BICePs score):
\begin{equation}
  \sigma_{\epsilon} = \sqrt{\mathrm{diag} \left[ \mathbf{H}^{-1} \right]}.
\label{eq:uncertainties}
\end{equation}
Conceptually, this means that the estimated uncertainties in the best-fit values represent the widths of the basins on the BICePs score landscape. This can sometimes lead to an overestimation of the uncertainty with trust-region optimization schemes, especially with flatter basins like those we observe in the BICePs score landscape. Conversely, the sharper the curvature (larger eigenvalues of the Hessian), the more certain we are about the location of the minimum in that direction, resulting in smaller computed uncertainties.  Inspection of Equations \eqref{eq:du}, \eqref{eq:d2f} and \eqref{eq:a16} show that $\sigma_{\epsilon} \propto 1/\sqrt{N}$, where $N$ is the number of BICePs replicas. This suggests that one can obtain smaller uncertainties in the optimized parameters by using more replicas.

\subsection*{Convergence criteria for SciPy optimization}
In all our multiparameter optimizations, we use a standardized convergence criterion, determined by assessing the average relative change between the old parameters $\epsilon^{\text{old}}$ and the new parameters $\epsilon_{i}^{\text{new}}$ at each iteration, calculated as
\begin{equation}
  \frac{1}{N_{\epsilon}} \sum_{i=1}^{N_{\epsilon}} \left| \frac{ \epsilon_{i}^{\text{old}} - \epsilon_{i}^{\text{new}} }{ \epsilon_{i}^{\text{old}} } \right| < 0.05,
  \label{eq:convergence_criteria}
\end{equation}
where convergence is considered to have been reached when the average change in parameters is less than 0.05.   In addition, we strictly require that the convergence condition is satisfied three times in succession.

\subsection*{Accuracy profiles}
Accuracy profiles quantify convergence of the optimized parameters $\vec{\epsilon}(t)$ toward the true parameters $\vec{\epsilon}^{*}$ as a function of optimization iteration $t$. We measure this using the relative $\ell_2$-error
\begin{equation}
\frac{\|\vec{\epsilon}(t) - \vec{\epsilon}^{*}\|_2}{\|\vec{\epsilon}^{*}\|_2} = \frac{\sqrt{\sum_i \left(\epsilon_i(t) - \epsilon_i^{*}\right)^2}}{\sqrt{\sum_i \left(\epsilon_i^{*}\right)^2}}.
  \label{eq:rel_L2_error}
\end{equation}
This metric provides a scale-invariant measure of parameter recovery, enabling direct comparison across optimization trajectories.

Uncertainty in this relative $\ell_2$-error arises from uncertainty in the estimated parameters $\vec{\epsilon}(t)$. At each iteration, parameter uncertainties are approximated using the diagonal of the inverse Hessian (Eq.~\ref{eq:uncertainties}), yielding standard deviations $\sigma_i(t)$ under a local Gaussian approximation of the objective landscape.

To propagate these uncertainties to the nonlinear quantity of Eq \ref{eq:rel_L2_error}, we use Monte Carlo error propagation. For each iteration $t$, we draw $N=5000$ samples $\epsilon_i^{(k)}(t) \sim \mathcal{N}\big(\epsilon_i(t), \sigma_i^2(t)\big)$, and compute the corresponding distribution of relative errors.  The uncertainty in the accuracy profile is then given by the sample standard deviation.  Final accuracy profiles and their uncertainties are reported as averages over 25 independent optimization runs.


\subsection*{Algorithms and settings for PyTorch optimization}
We represent all six $\{\epsilon_i\}$ parameters as PyTorch \texttt{nn.Parameter} objects and utilize automatic differentiation to compute gradients. Optimization is performed using the ADAM algorithm\cite{kingma2014adam} in conjunction with MCMC sampling over conformational states $X$ and uncertainty parameters $\sigma$, resulting in stochastic gradient estimates with non-negligible sampling variance. Empirically, a learning rate of $5 \times 10^{-2}$ yielded stable and rapid convergence; although larger than typical values used in deep neural network training, this reflects the low-dimensional parameter space and relatively small gradient magnitudes, with adaptive scaling mitigating instability. In contrast, optimization using SciPy's \texttt{minimize} internally determines appropriate step sizes based on local curvature information, making direct comparison of learning rate parameters between these approaches not meaningful. New trajectory data was generated every 20 steps, resulting in 100 training epochs, with convergence typically observed after approximately 20 epochs and minimal parameter updates thereafter.

\section{Results and Discussion}

\subsection*{The BICePs score landscape correctly identifies optimal parameters for an HP lattice model}

To understand BICePs' ability to learn  optimal parameters, we start by scanning the BICePs score landscape over a single $\epsilon$ parameter.  For this purpose, we use an HP model where all six hydrophobic beads share the same value of  $\{\epsilon_i\} = \epsilon$ .

To generate a set of experimental distance observables for testing, we set the "true"  value of $\epsilon$ to 1.0., and compute average distances and variances of eight distances for each macrostate (Figure \ref{fig:lattice_model}b).  The BICePs score and its first and second derivatives are then evaluated across a range of $\epsilon$ values, spanning from 0.0 to 5.5, in increments of 0.125, where prior energies are unique for each $\epsilon$ value. BICePs scores and derivatives are averaged over five independent calculations using 100k steps of MCMC with 8 replicas. The associated error bars represent the standard error of the mean across these five instances.

The resulting $f(\epsilon)$ landscape pinpoints the optimal parameter as a minimum, and correctly computes the first and second derivatives (Figure \ref{fig:score_landscape}). Despite the stochastic sampling inherent to BICePs, the BICePs score landscape is remarkably smooth and precise, revealing the optimal epsilon to be $\epsilon^{*} = 1.0$, the value with the lowest BICePs score, $f(\epsilon^{*})$ (Figure \ref{fig:score_landscape}a.), with its derivative $df/d\epsilon$ at this location equal to 0.0 (Figure \ref{fig:score_landscape}b.).  This optimal model is achieved when the overlap between the likelihood and prior distributions is maximized (Figure \ref{fig:flowchart}).    When $\epsilon$ is significantly larger than the true value, the uncertainty in $f$ and $\partial f/\partial \epsilon$ increases. In this case, the physical interpretation is that for large values of $\epsilon$, the 12-mer mainly populates the folded macrostate, which creates poor overlap between the likelihood and prior, resulting in more finite sampling error.

\begin{figure}
  \centering
  \includegraphics[width=\linewidth]{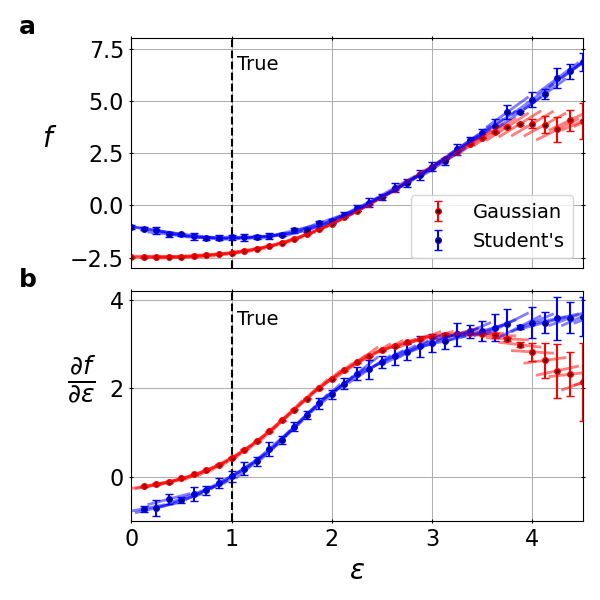}
  \caption{\small
      (a) 1-D scans over $\epsilon$ to reveal the landscape of the BICePs score (scatter dots). The tangent lines are the derivatives at each $\epsilon$ value. Uncertainties in the BICePs scores and derivatives come from the standard error of the mean across five independent scans along $\epsilon$. (b) The derivative of the BICePs score (scatter dots) and the second derivative of the BICePs score (tangent lines) at each epsilon value. The dotted black line at $\epsilon^{*}=1.0$ shows the true value, which is where the derivative of the BICePs score equals zero. BICePs calculations are run using the Gaussian likelihood model and the Student's likelihood model for 100k steps with 8 replicas. Systematic error (+4 L.U. shift) was induced to the 2-11 distance, with a total experimental error of $\sigma_{data}=1.41$. Unlike the Gaussian likelihood with $\left. \frac{\partial f}{\partial \epsilon} \!\right|_{\epsilon=1} = 0.43 \pm 0.002$, the Student's model accurately finds the global minima with $\left. \frac{\partial f}{\partial \epsilon} \!\right|_{\epsilon=1} = 0.02 \pm 0.07$.
    }
  \label{fig:score_landscape_with_error}
\end{figure}

\subsection*{The Student's model confers resiliency when dealing with experimental error}
We next examined the robustness of the BICePs score to errors and uncertainties present in experimental data, by introducing random and systematic errors to the 2--11 and 0--11 distance observables (Figure \ref{fig:score_landscape_with_error}).  When we used BICePs with a standard Gaussian likelihood model in which a single uncertainty parameter $\sigma$ is used for all experimental observables, introducing random and systematic errors significantly reduces the accuracy of the model.   When 2--11 and 0--11 distances are perturbed by +3 to +5 L.U. , the BICePs score predicts a value of $\epsilon$ smaller than the ``true'' value, predicting structures that are primarily "unfolded".  This in turn causes a leftward shift in the gradient plot (Figure \ref{fig:score_landscape_with_error}b).

In contrast, the Student's likelihood model deals with the outliers much more handily. The minimum BICePs score correctly identifies the optimal value of $\epsilon$, and the leftward shift in the gradient plot is absent.  The Student's model is able to automatically identify outliers directly from the input data, and mitigate their influence, even when 25\% of the data (2 of the 8 distances) are corrupted.

\begin{figure}
\centering
\includegraphics[width=\linewidth]{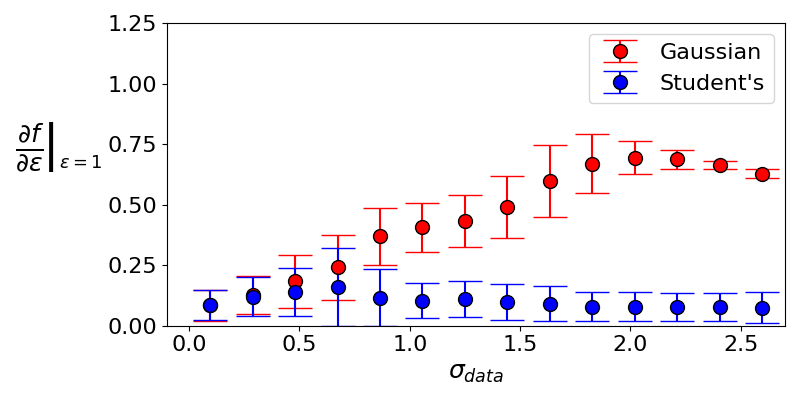}
  \caption{\small Comparative analysis in performance of the Student's likelihood model (blue) and a Gaussian likelihood model (red), when  random and systematic error of varying magnitude  ($\sigma_{data}$)  is introduced to the 2-11 and 0-11 distances.  The vertical axis shows the derivative of the BICePs score evaluated at the ``true'' target value,  $\epsilon=1$. This value corresponds to the "True" minima, illustrated in Figure \ref{fig:score_landscape_with_error}. The  Gaussian likelihood's derivative becomes notably less dependable when data incorporates errors, especially when $\sigma_{data}$ exceeds 0.75 L.U. surpassing one standard deviation. The values shown were calculated using 5000 random perturbations to the 2-11 and 0-11 distances, and represent the average of 300 BICePs calculations. Error bars represent the standard deviation.}
\label{fig:likelihood_comparison}
\end{figure}

To more critically assess the performance of the Gaussian and Student's models in dealing with systematic errors, we examine the extent to which the gradient of the BICePs score deviates from zero at the ``true'' value of $\epsilon^{*} = 1$. We introduced random perturbations to the 2--11 and 0--11 distances of various standard deviations, $\sigma_{data}$ (Figure \ref{fig:likelihood_comparison}). When the Gaussian likelihood model was used, the computed gradient at $\epsilon^{*} = 1$ increases proportionally with the magnitude of $\sigma_{data}$., and becomes particularly unreliable when the error surpasses 0.75 L.U. When the Student's model was used, however, we find that such deviations are reduced. As $\sigma_{data}$ increases from zero, the gradient at $\epsilon^{*} = 1$ begins to increase, but then begins to \textit{decrease} beyond $\sigma_{data}$ = 0.6 {L.U.}.  This happens because the perturbations in the distance observables become large enough to be detected as outliers by the Student's model.


\subsection*{Multiparameter optimization }

While our toy protein lattice model is very simple, the inference problem it presents is challenging, especially as we increase the number of epsilon parameters to refine.  Our 12-mer system has only eight experimental distances as experimental restraints, and these are significantly influenced by ensemble-averaging.  Moreover,  the protein lattice model has a highly cooperative folding landscape, which can lead to correlated distance restraints that don't independently provide informative insights.

With this in mind, we further tested multi-parameter minimization of the BICePs score, using the modified HP lattice model with tunable interaction strengths $\epsilon_i$. Ideally, our algorithm should be able to efficiently find the global minimum of $f(\epsilon_1, \epsilon_2, ...)$ starting from many different initial parameter sets,  and be resilient against random and systematic errors.  For two or more $\epsilon_i$ parameters, it becomes computationally expensive to perform multidimensional scans of the BICePs score landscape, necessitating automated parameter optimization.

We first generated test data for a model where we set the ``true'' values to  $\epsilon_2 = 1.25, \epsilon_4 = 1.5$ and the remaining $\epsilon_i$ set to 3.0. We then tested two-parameter  $(\epsilon_{2}, \epsilon_{4})$ optimizations, starting from three different initial parameter sets $(\epsilon_{2}^{0}, \epsilon_{4}^{0})$ = $\{(0.5, 5.0), (4.0, 5.0), (4.0, 0.5)\}$, using both L-BFGS-B and `trust-ncg' optimization for BICePs scores calculated using 8 replicas.

Optimizations using L-BFGS-B successfully converge very close to the ``true'' values of the interaction strength parameters ($\epsilon_{2}^{\text{True}}=1.25$, $\epsilon_{4}^{\text{True}}=1.5$) within 10 iterations (Figure \ref{fig:2-D_L-BFGS-B}).  We suspect upon tightening the convergence criteria (Eq. \ref{eq:convergence_criteria}) and running for 15-20 iterations, optimizations would get even closer to the ``true'' interaction strength parameters. The `trust-ncg' optimization, however, finds the correct minima in less than 10 iterations, demonstrating the notable advantage of second-order methods (Figure \ref{fig:2-D_trust-ncg}).  The landscape matches the computed BICePs scores, and shows that `trust-ncg' optimization trajectories of successive ($\epsilon_2, \epsilon_4$) values correctly follow gradients, normal to the landscape contour lines. Optimizations that used `trust-ncg' with 32 replicas resulted in average optimized parameters of ($\epsilon_2$ = 1.12 $\pm$ 0.42, and $\epsilon_4$ = 1.43 $\pm$ 0.36) (Figure \ref{fig:2-D_opt_epsilon_2_and_4_no_error}), which have much smaller uncertainty estimates compared to the calculations made with 8 BICePs replicas ($\epsilon_2$ = 1.07 $\pm$ 0.85, and $\epsilon_4$ = 1.46 $\pm$ 0.82).

\subsection*{Multiparameter optimization in the presence of systematic error}

To assess the resiliency of multiparameter optimization to systematic error,  we added extra complexity to the optimization problem by introducing systematic error of +3 and +3.5 L.U. to the 2--11 and 4--9 distances observables, respectively. This makes total error in the data $\sigma_{data} = 1.63$ L.U., a significant obstacle to the refinement of parameters $\epsilon_{2}$ and $\epsilon_{4}$, considering these beads each participate in only two of the measured distances  (Figure \ref{fig:lattice_model}b).

\begin{figure}
\centering
  \includegraphics[width=\linewidth]{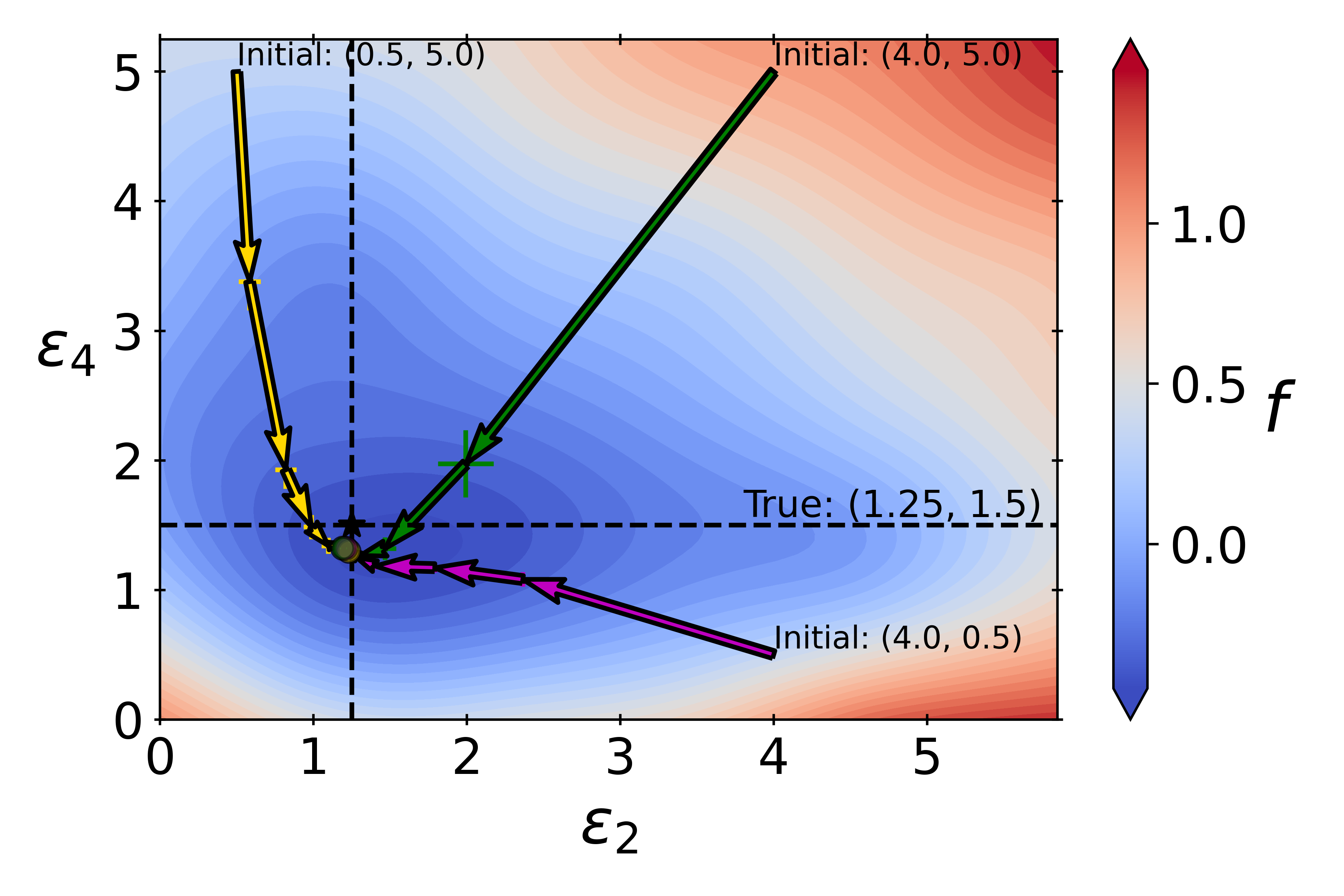}
  \caption{\small Average traces over a total of 25 independent rounds of parameter $(\epsilon_{2}, \epsilon_{4})$ optimizations using second-order (trust-ncg) method with BICePs, for a maximum of ten iterations. Optimizations converge to the same parameters ($\epsilon_{2}^{\text{True}}=1.25$, $\epsilon_{4}^{\text{True}}=1.5$) when starting from different initial parameters $(\epsilon_{2}^{0}, \epsilon_{4}^{0})$ = $\{(0.5, 5.0), (4.0, 5.0), (4.0, 0.5)\}$.   Average BICePs optimized parameters were determined to be
$\epsilon_2$ = 1.22 $\pm$ 0.44, and $\epsilon_4$ = 1.30 $\pm$ 0.36.,
  where the uncertainties are esimated from the inverse Hessian.  The BICePs score landscape was generated from the average values of five scans over $\epsilon_{2}$ and $\epsilon_{4}$. All calculations used the Student's model with 200k MCMC steps and 32 replicas. The experimental data is corrupted with systematic error in the 2-11 and 4-9 distances for +3 and +3.5 L.U. shift, respectively.  The total error in the data is $\sigma_{data} = 1.63$ L.U..
  }
  \label{fig:2-D_opt_epsilon_2_and_4}
\end{figure}

We performed `trust-ncg' optimization using the Student's model with 200k MCMC steps and 32 replicas, starting from three different initial parameters: $(\epsilon_{2}^{0}, \epsilon_{4}^{0})$ = $\{(0.5, 5.0), (4.0, 5.0), (4.0, 0.5)\}$ (Figure \ref{fig:2-D_opt_epsilon_2_and_4}).  Our `trust-ncg' optimizations with 32 BICePs replicas results in average optimized parameters of $\epsilon_2$ = 1.22 $\pm$ 0.44, and $\epsilon_4$ = 1.30 $\pm$ 0.36., Notably, this was achieved in only a few iterations, showing the algorithm's proficiency in finding the optimal parameters, irrespective of the initial reference parameters.

In adherence to best practices\cite{beiranvand2017best}, we compute accuracy profiles for two-parameter `trust-ncg' optimizations using 32 BICePs replicas with no systematic error (Figure \ref{fig:accuracy_profile_32_rep_epsilon_2_and_4_no_error}) and systematic error (Figure \ref{fig:accuracy_profile_32_rep_epsilon_2_and_4_systematic_error}).  Optimizations with no systematic error were repeated using only 8 BICePs replicas (Figure \ref{fig:accuracy_profile_8_rep_epsilon_2_and_4}). A comparison of the 8-replica and 32-replica results shows that the variance between independent optimization traces decreases as the number of replicas increases.  The error bars plotted in the accuracy profiles report uncertainties from Monte Carlo error propagation using the $\sigma_{\epsilon}$ estimated from the inverse Hessian.  In each case, it can be seen that the actual variation in the relative error across many optimizations is well within the Hessian-estimated uncertainty.

As we did with the one-parameter optimizations, we also examined the performance of two-parameter optimizations when using a standard Gaussian likelihood model with 8 replicas, in the absence of systematic error (Figure \ref{fig:2-D_landscape_systematic_error_Gaussian}a) and in the presence of systematic error (Figure \ref{fig:2-D_landscape_systematic_error_Gaussian}b). As before, we find that the accuracy of the refined parameters drops significantly when using a standard Gaussian likelihood model. With only a single parameter to estimate the uncertainty in the data set of distances, the Gaussian likelihood model is unable to properly detect the two outliers, and as a result predicts optimal $\epsilon_{2}$ and $\epsilon_{4}$ values that are much smaller than the ``true'' values. Conversely, the accuracy of the Student's model is very robust in the presence of systematic error (Figure \ref{fig:2-D_landscape_systematic_error_Students}).  These results yet again highlight the ability of BICePs to handle uncertainties and systematic error comprehensively when used with a likelihood model that can deal with outliers.


We next attempted to optimize a three-parameter BICePs score landscape, by introducing an additional parameter, $\epsilon_{6}$. This is a challenging task that may take many iterations to reach the convergence criteria. Our objective was to examine the typical percentage of optimization traces that converge within a designated number of iterations. For this test, we refrained from adding systematic error to the data.

\begin{figure}
\centering
  \includegraphics[width=\linewidth]{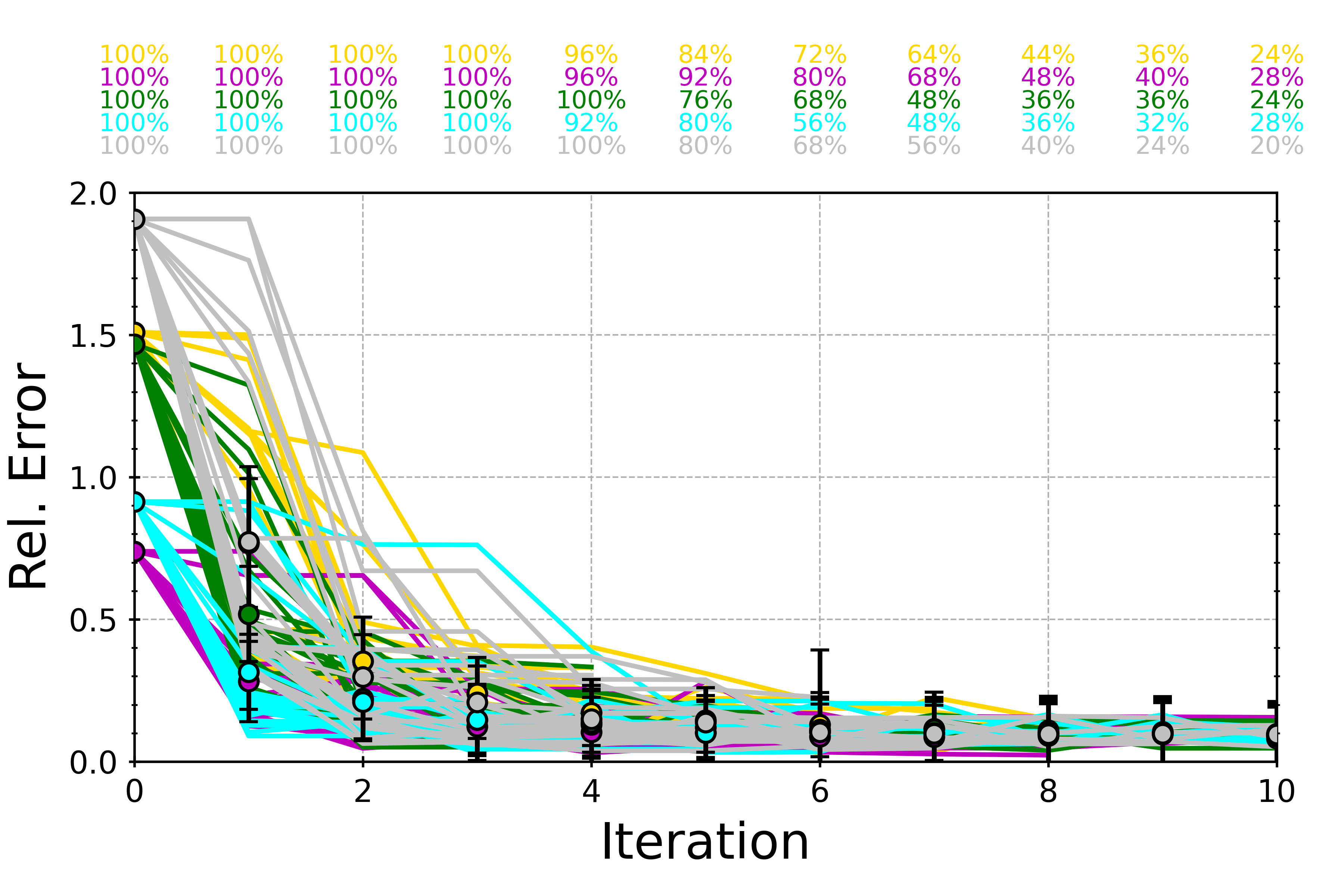}
  \caption{\small Accuracy profile of optimization traces, demonstrating convergence with a relative error of 0.1. A total of 25 independent rounds of parameter $(\epsilon_{2}, \epsilon_{4}, \epsilon_{6})$ optimizations were conducted for each of the five sets of initial parameters $(\epsilon_{2}^{0}, \epsilon_{4}^{0}, \epsilon_{6}^{0})$:
  $(4.0, 5.0, 3.0)$ (gray),
  $(0.5, 5.0, 0.5)$ (yellow),
  $(2.0, 4.0, 4.0)$ (green),
  $(1.0, 0.5, 3.0)$ (magenta),
  $(3.0, 0.5, 0.5)$ (cyan).
  Color-coded percentages report the fraction of trajectories that remain unconverged at each iteration. Error bars (black) represent the uncertainties in predicted parameters for each iteration, calculated using the diagonalized inverse Hessian (Equation \eqref{eq:uncertainties}) and averaged over the 25 independent optimizations. Average optimized parameter values from converged trajectories were $(\epsilon_{2}=1.11 \pm 0.42, \epsilon_{4}=1.37 \pm 0.35, \epsilon_{6}=1.47 \pm 0.38)$.
  }
  \label{fig:profile_epsilon_2_4_6}
\end{figure}

Five initial starting points were selected:  $(\epsilon_{2}^{0}, \epsilon_{4}^{0}, \epsilon_{6}^{0})$ = $(4.0, 5.0, 3.0),$ $(0.5, 5.0, 0.5)$, $(2.0, 4.0, 4.0)$, $(1.0, 0.5, 3.0)$, $(3.0, 0.5, 0.5)$. Optimization targeted the ``true'' interaction strength parameters: $\epsilon_{2}^{\text{True}}=1.25$, $\epsilon_{4}^{\text{True}}=1.5$, and $\epsilon_{6}^{\text{True}}=1.5$, with an acceptable average relative deviation of 0.1 (refer to Figure \ref{fig:profile_epsilon_2_4_6}).  Twenty-five independent optimizations were performed per starting point, using 32 BICePs replicas and 200k MCMC steps per iteration.

For these tests, we required that the convergence condition must be satisfied two times in succession, which is a typical stopping point.  Traces of the relative error throughout the optimization show a heterogeneous collection of convergence trajectories, depending on the starting point (Figure \ref{fig:profile_epsilon_2_4_6}). At each iteration, we track the percentage of traces that remain unconverged.
For some starting points, 5 out of the 25 optimization trajectories remain unconverged after ten iterations, while for others, up to 28\% remain unconverged going into the tenth iteration. Nevertheless, the accuracy profiles clearly demonstrate early convergence to the target. This indicates a mismatch between the convergence criterion and needless hovering around the optimal parameter values.  The average optimized parameters from the converged trajectories were determined to be $(\epsilon_{2}=1.11 \pm 0.42, \epsilon_{4}=1.37 \pm 0.35, \epsilon_{6}=1.47 \pm 0.38)$, where the uncertainties are calculated from the inverse Hessian. An overlay of average traces on the BICePs score landscape is shown in Figure \ref{fig:3-D_opt__epsilon_2_4_6}.

A robust correlation (coefficient of determination $R^{2}=0.83 \pm 0.14$) between the converged BICePs score and the relative error in the parameters further confirms the validity of our approach in higher dimensions (see Figure \ref{fig:3-D_sensitivity}). This relationship illustrates that an increased distance from the ``true'' parameters (i.e. a higher relative error) corresponds to an enlargement of the BICePs score $f$. A comparative analysis of average BICePs scores across optimization traces (Figure \ref{fig:3-D_score_vs_iteration}) and the landscape contours (Figure \ref{fig:3-D_opt__epsilon_2_4_6}) indicates that our Gaussian process regression successfully approximates the actual BICePs score landscape in higher dimensions.

\subsection*{Simultaneous refinement of all six interaction energy parameters with PyTorch integration}

To illustrate how our optimization approach naturally extends to larger neural network potentials, we simultaneously refine all six interaction energy parameters.  To do this, we integrated PyTorch into the BICePs framework, treating the parameters as trainable neural network weights and optimizing them via a combination of MCMC sampling and gradient-based updates. Previously, we showed how the BICePs score can serve as an objective function for training neural network forward models \cite{Raddi2024FMO}.

\begin{figure*}
\centering
  \includegraphics[width=\linewidth]{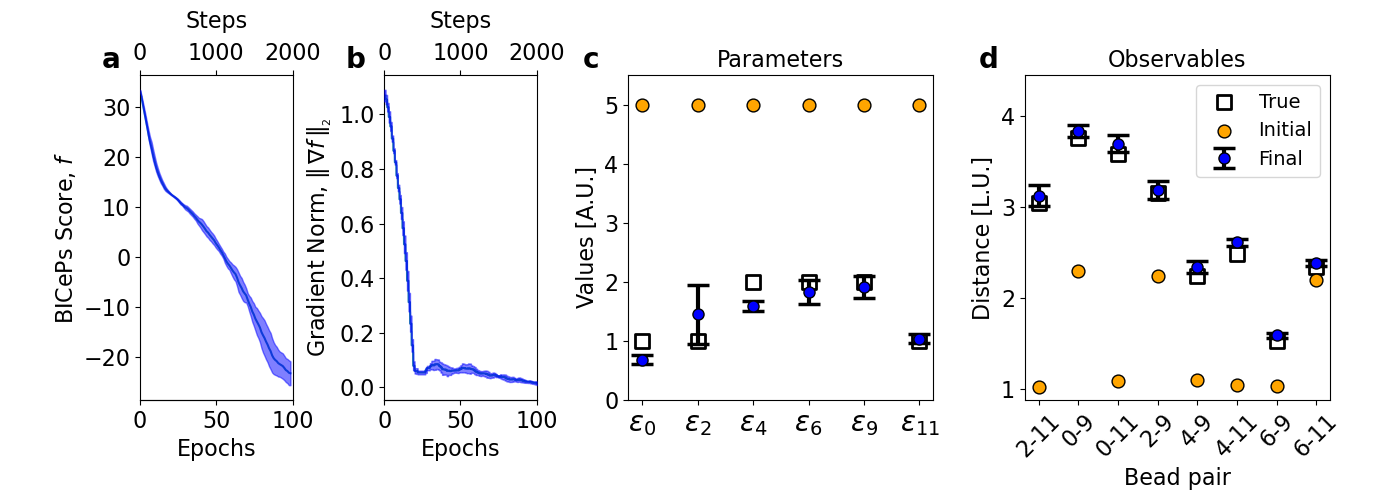}
  \caption{\small      Optimization of all six epsilon parameters using the Gaussian likelihood model. Results and uncertainties are taken as the average and standard deviation across five independent optimizations. (a) The BICePs score $f$ across training iterations. (b) The gradient norm of the BICePs score converges toward zero. (c) Optimized $\epsilon$ values closely overlap the true interaction energy paramters. The prior model contained initial parameters all set to $\epsilon^{0}=5.0$.  The true $\epsilon$ parameters were ($\epsilon_{0}^{*}=1.0$, $\epsilon_{2}^{*}=1.0$, $\epsilon_{4}^{*}=2.0$, $\epsilon_{6}^{*}=2.0$, $\epsilon_{9}^{*}=2.0$, $\epsilon_{11}^{*}=1.0$), while the optimized parameters were determined to be ($\epsilon_{0}=0.68 \pm 0.07$, $\epsilon_{2}=1.45 \pm 0.50$, $\epsilon_{4}=1.59 \pm 0.09$, $\epsilon_{6}=1.83\pm 0.20$, $\epsilon_{9}=1.92 \pm 0.19$, $\epsilon_{11}=1.04 \pm 0.07$), yielding a parameter mean absolute error (MAE) of 0.12.  (d) Comparison of the ensemble average distance observables. The distances computed using the optimized parameters gave a MAE of 0.080 L.U. to the true distances. Here, no random or systematic error was placed in the experimental data.
  }
  \label{fig:all_epsilons}
\end{figure*}

All parameters were initialized far from their true values at $\epsilon_{i}^0 = 5.0$, and synthetic experimental data was generated without random or systematic error. As shown in Figure \ref{fig:all_epsilons}, the BICePs score steadily improved over training iterations and looks to continue decreasing due to sigma values getting very small (Figure \ref{fig:all_epsilons} a), with the gradient norm (Figure \ref{fig:all_epsilons} b) decreasing toward convergence. All results shown in Figure \ref{fig:all_epsilons} are taken as average and standard deviation across five independent optimizations.  The optimized ensemble accurately recovered the true pairwise distances with a mean absolute error (MAE) of 0.080 L.U. (Figure \ref{fig:all_epsilons} c). The optimized parameters were determined to be ($\epsilon_{0}=0.68 \pm 0.07$, $\epsilon_{2}=1.45 \pm 0.50$, $\epsilon_{4}=1.59 \pm 0.09$, $\epsilon_{6}=1.83\pm 0.20$, $\epsilon_{9}=1.92 \pm 0.19$, $\epsilon_{11}=1.04 \pm 0.07$), yielding a parameter MAE of 0.12 (Figure \ref{fig:all_epsilons} d).

These findings demonstrate that BICePs, augmented with PyTorch, can efficiently and accurately recover interaction parameters and ensemble observables through simultaneous multi-parameter refinement. The seamless integration of gradient-based optimization with Bayesian sampling provides a robust and scalable approach, paving the way for training complex neural network potentials in molecular modeling applications.


\subsection*{Application to the toy polymer model of K\"{o}finger \& Hummer}

K\"{o}finger and Hummer (KH) recently developed a method to optimize molecular simulation force fields using a Bayesian inference maximum-entropy reweighting method called BioFF.\cite{kofinger2021empirical}  To validate their method, they introduced a simple parameterizable 2-D polymer model, and demonstrated that BioFF can efficiently discover the optimal parameters that reproduce ensemble-averaged observables. Here, we demonstrate that BICePs optimization can also reproduce this result.

The KH polymer model is a chain of $N$ beads, each separated by unit length, with orientations defined by $N-1$ angles $\Delta \phi$ between contiguous bond vectors. Each bond angle is distributed according to a von Mises distribution parameterized by $\kappa$, which controls the polymer's stiffness. Full details about this model are given in Appendix \ref{sec:polymer_model}.  Unlike the HP lattice model, which possesses an exact solution for the prior, the KH polymer model requires estimation of the prior through statistical sampling, which is subject to finite sampling error.

To implement KH model optimization with BICePs, we derived the first and second partial derivatives of the energy function required for computation of the BICePs score (see SI).  Following K\"{o}finger and Hummer,\cite{kofinger2021empirical} we consider a 100-bead chain, and use the end-to-middle distance (between bead 1 and bead 50) as the ensemble-averaged observable, which we estimate as the mean of 10000 samples from the von Mises distribution with the ``true'' value of $\kappa$ set to 10.0.

We then performed `trust-ncg' optimization using 200 BICePs replicas and 100k MCMC steps, starting from an initial parameter value of $\kappa$ = 20.0.  At the first optimization step, draw 500 samples from the von Mises distribution, and use their energies to compute their Boltzmann populations; each sample is a separate conformational state. In each subsequent optimization step, we draw 500 more samples to add to the collection, and compute energies and Boltzmann populations for the combined collection. The result of this procedure is highly efficient discovery of the optimal value of $\kappa^{*}$ = 10, reaching convergence within two iterations (Figure \ref{fig:polymer_model}).

While BICePs optimization achieves similar results to BioFF, it is worth mentioning some of the key differences of the two methods.  BioFF is a maximum-entropy method that reweights a set of reference populations by minimizing a negative log-likelihood that uses a regularization parameter, $\theta$, to balance two terms: the relative entropy with respect to the reference, and a $\chi^2$ term representing the deviation between model predictions of ensemble-averaged quantities and experiment. The BioFF-based optimization algorithm iteratively (1) reweights the populations given a current set of force field parameters, then (2) optimizes the force field parameters given the updated populations, repeating this cycle until convergence.

One issue with this approach is that it requires a coarse-grained description of the relevant conformational states and their populations, which need to be revised as the force field parameters are optimized. K\"{o}finger and Hummer deal with this by keeping track of each new thermodynamic ensemble being sampled, and using the running collection of all the samples obtained to estimate each new set of coarse-grained state populations with the binless WHAM\cite{Shirts:2008eza,rosta2011catalytic} free energy estimator.   In our BICePs approach, we avoid this expense by simply adding new samples as additional conformational states, and recalculating populations as their Boltzmann weights.

Another issue with maximum-entropy methods like BioFF\cite{kofinger2019inferring,kofinger2021empirical} and BioEn\cite{bottaro2020integrating} is the need to choose an appropriate value for the regularization parameter $\theta$. BICePs does not require a regulation parameter, because the BICePs score is computed by sampling the entire posterior distribution of uncertainty parameters $\sigma$.  (In BICePs, the $\sigma$ are the analogous quantities that balance prior information against the experimental restraints.)



Very recently, K\"{o}finger \& Hummer put the regularization parameter $\theta$ on a firmer theoretical footing by relating it to the variance of the reduced energy difference, thus enabling estimation of $\theta$ \textit{a priori}.\cite{kofinger2023encoding}. Another recent work, Gilardoni et al. \cite{gilardoni2023boosting} introduced an approach similar to BICePs, in that ensemble refinement and force field parameter refinement are integrated into a singular process. This method bears resemblance to both BioEn and BioFF, as it employs a regularization parameter to scale the Kullback-Leibler divergence. Gilardoni et al.'s technique is distinguished by the incorporation of two distinct confidence hyperparameters. The first parameter is designed to adjust the reliability of the prior, while the second modulates the trustworthiness of the force field corrections.

\subsection*{Further Discussion}

Our work here greatly improves upon previous work by Ge et al., which used an HP lattice model as an example system to perform force field parameterization and model selection using BICePs.\cite{ge2018model}.  That work approached the problem through brute force, scanning parameter space in search of the minimum BICePs score, which would be costly and impractical for multiple parameters.  Another issue with that work was the use of a previous version of BICePs (called BICePs v1.0) that does not use replica-averaging, and therefore was not a maximum-entropy method.  Because of this, applying BICePs in practical cases, such as refining conformational populations against NMR observables for beta-hairpins trploop 2A and 2B, resulted in ensembles dominated by the state(s) that best agree with experiment.  Since that work, we have developed the current version of BICePs that performs replica averaging and is equipped with specialized likelihood functions that can account for systematic error.\cite{raddi2023biceps} The HP lattice model continues to serve as an excellent test system for force field refinement as it allows us to use multiple parameters and provides exact solutions for generated ensembles.


While our algorithm has the advantage of being able to incorporate priors on force field parameters, this is a possibility we do not pursue in this work.  Obtaining priors for force field parameters is a more involved approach, but we leave room for this in future development.

\subsubsection*{Computational cost}

Given the extra overhead of posterior sampling required by BICePs compared to other methods, it is important to discuss the computational cost of BICePs optimization protocols, and how this cost might scale with the number of parameters involved in the optimization.

First, we note that the computational cost of BICePs or any other force field optimization scheme seeking to optimize parameters against ensemble-averaged measurements is dominated by the expense of having to perform enough conformational sampling to obtain statistically reliable estimates of thermodynamic averages.  This will be highly system-dependent, but can be considered a fixed cost per iteration.  BICePs treats uncertainty parameters as random variables and samples over them during optimization. This introduces additional computational cost, but importantly, because the dimensionality of the nuisance space is typically low compared to the effective dimensionality of the system, the associated overhead is minimal, and contributes to a slightly larger fixed cost per iteration.

Next, we consider how the computational expense scales with the number of observables, $m$, that need to be computed, and the number of parameters $n$ that need to be optimized. This expense will scale with $m$ and $n$ in the same way for BICePs as it would for other objective functions (e.g. minimization of a $\chi^2$ value).  At each iteration, $m$ observables, calculated as thermodynamic averages, need to be computed and stored, regardless of the optimization method, resulting in an $\mathcal{O}(m)$ cost.  For gradient-based optimization, the additional thermodynamic averages of $n$ first-order derivatives (the Jacobian) must be computed and stored; these quantities have terms that must be computed for every observable, resulting in an $\mathcal{O}(m \times n)$ cost. For Hessian-based optimization, $n(n-1)/2$ thermodynamic averages of second-order derivatives (the Hessian) must also be computed and stored, resulting in a $\mathcal{O}(m \times n^2)$ cost.
To demonstrate this scaling empirically, we evaluated runtimes of our BICePs optimization code applied to the HP lattice model system with different numbers (2, 3, 4, and 5) of epsilon parameters (Figure \ref{fig:runtime_eval}).  As expected, we found that increasing the parameter count from n=2 to n=5 results in negligible change in sampling time, approximately linear growth in the Jacobian evaluation (0.55 s to 1.99 s), and approximately quadratic growth in the Hessian evaluation (0.93 s to 6.55 s).

We note that occasional convergence issues in MBAR (e.g., failure to reach tolerance with adaptive solvers) can introduce minor variability in runtime, but these cases are infrequent and do not significantly impact overall performance. All reported runtimes were evaluated using a MacBook M1 Pro.

Optimization of all six parameters over 100 epochs using BICePs sampling with first-order minimization implemented in PyTorch requires approximately 18.9 seconds. However, this approach is much faster due to derivatives being computed on-the-fly after every 20 steps of sampling $X$ and $\sigma$.

\subsubsection*{Comparing the computation cost to a simpler objective function such as a $\chi^2$.}

An alternative to BICePs is to minimize a less rigorous objective function such as a $\chi^2$ metric, where $\chi^2 = \sum_j \big( (d_j - f_j(\mathbf{X}))/\sigma_j\big)^2$. The computational cost of evaluating the objective scales linearly with the number of observables, making it appealing from a scaling perspective.   However, there are several reasons to believe the improved accuracy of BICePs is worth the expense.   The first is that the extra expense is minimal. BICePs does not assume values for the $\sigma_j$, and instead infers them by treating the uncertainty parameters as random variables and sampling over them during optimization. Importantly, because the dimensionality of this nuisance space is relatively low (in practice many observables are coupled to a single $\sigma$ value), the associated overhead is modest. As a result, the overall computational scaling of BICePs is not governed by uncertainty sampling, but rather by the cost of derivative evaluations with respect to the force field parameters.

Another reason is that even though a $\chi^2$-based approach scales well, some principled procedure needs to be used to determine the optimal fixed values of $\sigma_j$, which incurs some additional computational cost.  In their Bayesian MaxEnt approach to ensemble reweighting, K\"{o}finger and Hummer established an elegant theoretical connection between the regularization parameter $\theta$ and the variance of reduced free energy differences, enabling its estimation \textit{a priori}. Prior to this development, $\theta$ was typically selected by scanning an L-shaped curve, effectively introducing an outer optimization loop. If such a procedure were incorporated into each iteration of force field optimization, the resulting computational cost would scale multiplicatively with the number of candidate $\theta$ values, rendering it comparable to—or potentially exceeding—that of the BICePs approach.

Taken together, we believe these considerations suggest that when uncertainty is handled consistently, the cost difference between $\chi^2$-based approaches and BICePs becomes marginal, while the gains in robustness and accuracy remain substantial.

\subsection*{Optimization across multiple systems}
An eventual application of our BICePs optimization method is transferable force field parameterization across many different molecular systems (e.g. a set of proteins) against ensemble-averaged experimental measurements.

Our approach can easily be adapted for this purpose, and performed efficiently so that each molecular system can be evaluated in a separate BICePs calculation.  To do this, suppose we have a set of molecular systems indexed by $s$, of total number $N_s$. We make use of the partial derivatives of the BICePs score with respect to each $\epsilon_i$ to determine a weight for each system $s$.  The normalized weight for the $i^{th}$ parameter of system $s$ is
\begin{equation}
  w_{i,s} = \frac{ \left( \frac{\partial f}{\partial \epsilon_{i} } \right)_{s} \left( \epsilon_{i,s} - \epsilon_{i}^{\text{old}} \right) }{ \sum_{s=1}^{N_{s}}\left( \frac{\partial f}{\partial \epsilon_{i} } \right)_{s} \left( \epsilon_{i,s} - \epsilon_{i}^{\text{old}} \right)}.
\end{equation}
At each optimization iteration, we update the $i^{th}$ parameter, $\epsilon_{i}^{\text{new}}$ according to a weighted average across all of the systems:
\begin{equation}
  \epsilon_{i}^{\text{new}} \leftarrow \sum^{N_{s}}_{s=1} w_{i,s} \epsilon_{i,s}.
\end{equation}
In this scheme, the derivatives play a pivotal role, where the weights are determined by how significantly each parameter change influences the BICePs score. This provides a balance that ensures that the parameters of a system that exert more effects on the score have more influence towards determining the next set of parameters, $\epsilon_{\text{new}}$.

\subsection*{Outlook}
Looking ahead, there is tremendous potential to extend the scope of the BICePs optimization against ensemble averaged observables to many applications.   An area of application would be to optimize physics-based molecular mechanics force fields for proteins against the many published solution-NMR data that exist.\cite{cavender2023structure}  Another  application would be to optimize parameters of general-purpose force fields for bespoke applications.  Examples of this may include refining existing all-atom microscopic models to better predict macroscopic properties (e.g. ensemble-average properties of disordered peptides like radius of gyration), or developing classes of models that can better predict the experimental observables for peptidomimetics without having to perform expensive fitting to quantum mechanical energy surfaces. In principle, any parameterizable model for the prior can be used with BICePs, including neural-network based potentials, or even generative models obtained from deep learning methods.\cite{lewis2025scalable,jing2025aibasedmethodssimulatingsampling}

\section{Conclusion}

In this work, we showed how the BICePs approach can be used for automatic and robust optimization of force field parameters against ensemble-averaged experimental measurements.  This is done by variational minimization of the BICePs score, a free energy-like quantity characterizing how well a model agrees with the experimental data, calculated by sampling over the posterior distribution of possible uncertainties. Using information from first and second partial derivatives of the BICePs score with respect to model parameters, we show how robust multi-parameter optimization of the BICePs score can be performed. Demonstrations in simple toy protein lattice models and polymer models show the utility of the method, and how specialized likelihood functions to deal with outliers are especially good at dealing systematic error. These results suggest automated variational optimization of the BICePs score is a powerful approach for the parameterization of molecular potentials.

\section*{Conflicts of interest}
Authors declare no conflicts of interest.

\section*{Data and software}
The BICePs algorithm is openly available at \href{https://github.com/vvoelz/biceps}{github.com/vvoelz/biceps}. All calculations in this work were performed using the development version biceps\_v3.0a, which is planned to be merged into the main branch. The specific version can be accessed at: \href{https://github.com/vvoelz/biceps/tree/biceps_v3.0a}{github.com/vvoelz/biceps/tree/biceps\_v3.0a}. The HP lattice model is open source software, located at:
\href{https://github.com/vvoelz/HPSandbox}{github.com/vvoelz/HPSandbox}.  For any issues or questions, please submit the request on GitHub.

\section*{Acknowledgements}
  The authors acknowledge that substantial content in this manuscript has also been included in the doctoral dissertation of RMR.\cite{raddi2024thesis} RMR and VAV are supported by National Institutes of Health grant R01GM123296.

\appendix

\label{sec:SI_theory}

\section{Explicit evaluation of marginalization integral to yield the lower incomplete gamma function inside the Student's likelihood.}
\label{sec:marginal_integral}
Here, we explicitly show how we arrive at the right hand side of equation \ref{eq:posterior_students} after marginalization over $\sigma_{j}$.
First, we collect all terms containing $\sigma_j$, then the integral takes the form
\begin{equation}
\int\limits_{\sigma^{\mathrm{SEM}}}^{\infty} \sigma_j^{-(2\beta+1)} \exp\!\left[   -\frac{\left(d_j - f_j(\mathbf{X})\right)^{2} + 2\beta\sigma_0^{2}}{2\sigma_j^{2}} \right] \, d\sigma_j ,
\end{equation}
up to multiplicative constants independent of $\sigma_j$.

We now perform the change of variables
\begin{equation}
  \begin{aligned}
    u &\equiv \frac{\left(d_j - f_j(\mathbf{X})\right)^{2} + 2\beta\sigma_0^{2}}{2\sigma_j^{2}}\\
    d\sigma_j &= -\frac{1}{2} \left(\frac{\left(d_j - f_j(\mathbf{X})\right)^{2} + 2\beta\sigma_0^{2}}{2}\right)^{1/2} u^{-3/2} \, du .
  \end{aligned}
\nonumber
\end{equation}

The integral therefore reduces to a constant $A$ multiplied by the standard definition of the lower incomplete gamma function,
\begin{equation}
  \begin{aligned}
    A \int_{0}^{\frac{\left(d_j - f_j(\mathbf{X})\right)^{2} + 2\beta\sigma_0^{2}}{2(\sigma^{\mathrm{SEM}})^2}}& u^{\beta-1} e^{-u}\, du \\
    &= A \, \gamma\!\left( \beta, \frac{\left(d_j - f_j(\mathbf{X})\right)^{2} + 2\beta\sigma_0^{2}}{2(\sigma^{\mathrm{SEM}})^2} \right),
  \end{aligned}
\end{equation}
where
\begin{equation}
A = \frac{1}{2}\left[ \beta \sigma_0^2\left(1+\frac{\left(d_j-f_j(\mathbf{X})\right)^2}{2 \beta \sigma_0^2}\right)\right]^{-\beta}
\nonumber
\end{equation}

Combining this result with the remaining normalization constants yields the likelihood expression on the right hand side of equation \ref{eq:posterior_students}.

\section{Deriving the first and second derivatives of the BICePs score.}
The BICePs score is defined as the negative logarithm of the Bayes factor, which is the quotient of two model evidences:
\begin{equation}
  f(\epsilon) = - \ln \left(Z(\epsilon) \big/ Z_{0}\right) ,
\end{equation}
where the evidence for model with force field parameters $\epsilon$ is
\begin{equation}
  Z(\epsilon) = \int  \int   \exp\left(-u(\mathbf{X}, \mathbf{\sigma} \mid D, \epsilon)\right) d \mathbf{X} d \mathbf{\sigma}
\end{equation}
and $Z_{0}$ is the model evidence for the reference distribution, which is set to be a uniform ensemble (see Figure \ref{fig:flowchart}) and so, does not depend on $\epsilon$.  The partial derivatives of the BICePs score, $f(\epsilon)$ , with respect to the model parameters  $\epsilon_i$ , is:
\begin{equation}
  \frac{\partial f(\epsilon)}{\partial \epsilon_{i}}=\frac{\partial}{\partial \epsilon_{i}}\left[-\ln \frac{Z(\epsilon)}{Z_0}\right]=-\frac{1}{Z} \frac{\partial Z(\epsilon)}{\partial \epsilon_{i}},
\end{equation}
where
\begin{equation}
\frac{\partial Z}{\partial \epsilon_{i}}=\int\int\left[-\frac{\partial u}{\partial \epsilon_{i}}\right] \exp (-u) d X d \sigma
  \label{eq:dZ}
\end{equation}
This allows us to express the derivative of the BICePs score with respect to the force field parameter $\epsilon_{i}$  as:
\begin{equation}
\begin{split}
  \frac{\partial f(\epsilon)}{\partial \epsilon_{i}} &= \int \int \frac{1}{Z(\epsilon)}  \left[ \frac{\partial u}{\partial \epsilon_{i}} \right] \exp \left( - u \right)  d \mathbf{X} d \mathbf{\sigma}\\
\end{split}
\end{equation}
which is the difference of Boltzmann averaged values of $\partial u / \partial \epsilon_{i}$.  This can also be explained in terms of prior energies. BICePs uses $\lambda$ to scale the prior energies, where the reference ensemble is given $\lambda=0$. Therefore, the average energy of the reference ensemble is zero and
\begin{equation}
\begin{split}
  \frac{\partial f(\epsilon)}{\partial \epsilon_{i}} = \bigg\langle \frac{\partial u}{\partial \epsilon_{i}} \bigg\rangle
\end{split}
\end{equation}

The Hessian matrix and corresponding second partial derivatives of the BICePs score is  required for computing uncertainties of the inferred parameters by the covariance matrix (see equation \ref{eq:uncertainties}), and when using a trust-region methods for optimization (see methods like `trust-ncg`, `trust-exact` in `scipy.optimize.minimize`). Trust-region methods can be more robust for noisy gradients; they may perform better with noisy objective functions because it does not rely solely on the gradient information to determine the search direction. Instead, it uses a combination of the gradient and second-order information with the local quadratic model to estimate the direction of steepest descent.  The Hessian matrix contains elements of second-order partial derivatives with respect to each parameter and gives the curvature of the objective function.
\begin{equation}
\begin{aligned}
  \mathbf{H} = \begin{bmatrix}
\frac{\partial^2 f}{\partial \epsilon_1^2} & \frac{\partial^2 f}{\partial \epsilon_1\partial \epsilon_2} & \cdots & \frac{\partial^2 f}{\partial \epsilon_1\partial \epsilon_n}\\
\frac{\partial^2 f}{\partial \epsilon_2\partial \epsilon_1} & \frac{\partial^2 f}{\partial \epsilon_2^2} & \cdots & \frac{\partial^2 f}{\partial \epsilon_2\partial \epsilon_n}\\
\vdots & \vdots & \ddots & \vdots \\
\frac{\partial^2 f}{\partial \epsilon_n\partial \epsilon_1} & \frac{\partial^2 f}{\partial \epsilon_n\partial \epsilon_2} & \cdots & \frac{\partial^2 f}{\partial \epsilon_n^2}
\end{bmatrix}
\end{aligned}
\end{equation}
The second-order partial derivative of the BICePs score for parameters $i$ and $j$, where $i \neq j$ is given as
\begin{equation}
  \label{eq:d2f_ij}
  \begin{split}
    \frac{\partial^2 f}{\partial \epsilon_{i} \partial \epsilon_{j}}&=\left[-\frac{1}{Z(\epsilon)} \frac{\partial^2 Z(\epsilon)}{\partial \epsilon_{i} \partial \epsilon_{j}}+\frac{1}{Z(\epsilon)^{2}} \frac{\partial Z(\epsilon)}{\partial \epsilon_{i}} \cdot \frac{\partial Z(\epsilon)}{\partial \epsilon_{j}} \right],
  \end{split}
\end{equation}
where
\begin{equation}
  \begin{split}
    \frac{1}{Z(\epsilon)}\frac{\partial^2 Z}{\partial \epsilon_{i} \partial \epsilon_{j}} &= \frac{1}{Z}\frac{\partial }{\partial \epsilon_{j}} \left[  \iint \left(- \frac{\partial u}{\partial \epsilon_{i}} \right) \exp(-u) dX d\sigma \right] \\
    & = \frac{1}{Z}\iint \left[\frac{\partial u}{\partial \epsilon_{i}} \cdot \frac{\partial u}{\partial \epsilon_{j}}-\frac{\partial^2 u}{\partial \epsilon_{i} \partial \epsilon_{j}}\right] \exp \left(-u\right) d X d \sigma \\
    &= \bigg\langle  \frac{\partial u }{\partial \epsilon_{i}} \cdot \frac{\partial u }{\partial \epsilon_{j}} \bigg\rangle -\bigg\langle \frac{\partial^{2} u }{\partial \epsilon_{i}\partial \epsilon_{j}} \bigg\rangle
  \end{split}
\end{equation}
and plugging equation \ref{eq:dZ} into the second term of equation \ref{eq:d2f_ij} gives
\begin{equation}
  \begin{split}
    \frac{1}{Z(\epsilon)^{2}} \frac{\partial Z}{\partial \epsilon_i} \frac{\partial Z}{\partial \epsilon_j} &= \left[ \frac{1}{Z} \iint \frac{\partial u}{\partial \epsilon_i} e^{-u} \, dX \, d\boldsymbol\sigma \right] \left[ \frac{1}{Z} \iint \frac{\partial u}{\partial \epsilon_j} e^{-u} \, d X \, d\boldsymbol\sigma\right] \\
    &= \left\langle \frac{\partial u}{\partial \epsilon_i} \right\rangle \left\langle \frac{\partial u}{\partial \epsilon_j}\right\rangle .
  \end{split}
\end{equation}
After substitution, equation \ref{eq:d2f_ij} becomes
\begin{equation}
  \begin{aligned}
    \frac{\partial^{2} f(\epsilon)}{\partial \epsilon_{i} \partial \epsilon_{j}} &= \bigg\langle \frac{\partial^{2} u }{\partial \epsilon_{i}\partial \epsilon_{j}} \bigg\rangle
    - \bigg\langle  \frac{\partial u }{\partial \epsilon_{i}} \cdot \frac{\partial u }{\partial \epsilon_{j}} \bigg\rangle
    +\bigg\langle \frac{\partial u }{\partial \epsilon_{i}} \bigg\rangle \bigg\langle\frac{\partial u }{\partial \epsilon_{j}} \bigg\rangle \\
    &= \left\langle \frac{\partial^{2} u}{\partial \epsilon_{i}\partial \epsilon_{j}} \right\rangle - \operatorname{Cov} \!\left( \frac{\partial u}{\partial \epsilon_i}, \frac{\partial u}{\partial \epsilon_j}\right) .
  \end{aligned}
\end{equation}

When $i = j$ (diagonal elements of the Hessian), the second-order partial derivatives of the BICePs score can be seen as the difference of the Boltzmann-averaged second-order partial derivative of the energy and the variance in its first-order partial derivative:
\begin{equation}
  \begin{split}
    \frac{\partial^{2} f(\epsilon)}{\partial \epsilon^{2}} = \bigg\langle \frac{\partial^{2} u }{\partial \epsilon^{2}} \bigg\rangle
    - \left( \bigg\langle \left( \frac{\partial u }{\partial \epsilon}\right)^{2} \bigg\rangle
    -\bigg\langle \frac{\partial u }{\partial \epsilon} \bigg\rangle^{2} \right)
  \end{split}
\end{equation}

To compute the derivatives of the BICePs score, we need the first and second derivatives of the BICePs energy function $u$, which consists of taking derivatives of the negative logarithm of the posterior, since $u = -\log p(X,\boldsymbol{\sigma}|D, \epsilon)$. The derivative of this function with respect to some force field parameter $\epsilon$ is

\begin{equation}
\begin{split}
  \frac{\partial u}{\partial \epsilon_{i}}  &=\frac{\partial E}{\partial \epsilon_i} \left( \sum^{N}_{r=1} -\log \left[ \frac{\exp (-E(X_{r} | \epsilon))}{\sum_{X} \exp \left( -E(X | \epsilon)\right)}\right] \right)\\
  &= - \sum^{N}_{r=1} \frac{\sum_{X} \exp \left( -E(X | \epsilon)\right)}{\exp \left( -E(X | \epsilon)\right)} \frac{\partial}{\partial \epsilon_{i}} \left[ - \frac{\frac{\partial E(X|\epsilon)}{\partial \epsilon_{i}} \exp (-E(X | \epsilon))}{\sum_{X} \exp \left( -E(X | \epsilon)\right)} \right.\\
  &\left. \quad\quad- \frac{\sum_{X} -\exp \left( -E(X | \epsilon)\right) \frac{\partial E(X|\epsilon)}{\partial \epsilon_{i}} \exp (-E(X | \epsilon))}{\left(\sum_{X} \exp \left( -E(X | \epsilon)\right) \right)^{2}} \right]\\
&= \sum^{N}_{r=1} \biggl\{ \frac{\partial E(X_{r}|\epsilon)}{\partial \epsilon_{i}} - \frac{\sum_{X} \exp \left( -E(X | \epsilon)\right) \frac{\partial E(X|\epsilon)}{\partial \epsilon_{i}}}{\sum_{X} \exp \left( -E(X | \epsilon)\right)} \biggr\}\\
  &= \sum^{N}_{r=1} \biggl\{ \frac{\partial E(X_{r}|\epsilon)}{\partial \epsilon_{i}} - \bigg\langle \frac{\partial E(X|\epsilon)}{\partial \epsilon_{i}} \bigg\rangle \biggr\}
\end{split}
\end{equation}

The second derivative is
\begin{equation}
\frac{\partial^2 u}{\partial \epsilon_i \partial \epsilon_j} = \sum_{r=1}^N \left[ \frac{\partial^2 E(X_r|\epsilon)}{\partial \epsilon_i \partial \epsilon_j} - \frac{\partial}{\partial \epsilon_j} \bigg\langle \frac{\partial E(X|\epsilon)}{\partial \epsilon_i} \bigg\rangle\right],
  \label{eq:d2u}
\end{equation}
where
\begin{equation}
\begin{split}
  \frac{\partial}{\partial \epsilon_j} \bigg\langle \frac{\partial E}{\partial \epsilon_i} \bigg\rangle &= \bigg\langle \frac{\partial^2 E}{\partial \epsilon_i \partial \epsilon_j} \bigg\rangle - \bigg\langle \frac{\partial E}{\partial \epsilon_i} \frac{\partial E}{\partial \epsilon_j} \bigg\rangle + \bigg\langle \frac{\partial E}{\partial \epsilon_i} \bigg\rangle \bigg\langle \frac{\partial E}{\partial \epsilon_j} \bigg\rangle.
\end{split}
\end{equation}
After substituting the derivative of the Boltzmann average into equation \ref{eq:d2u}, the second derivative becomes
\begin{equation}
  \begin{split}
    \frac{\partial^{2} u }{\partial \epsilon_{i}\partial \epsilon_{j}} &= \sum^{N}_{r=1} \biggl\{ \frac{\partial^{2} E(X_{r} | \epsilon)}{\partial \epsilon_{i}\partial \epsilon_{j}}
    - \bigg\langle \frac{\partial^{2} E}{\partial \epsilon_{i}\partial \epsilon_{j}} \bigg\rangle\\
    &\quad\quad\quad+ \bigg\langle \frac{\partial E}{\partial \epsilon_{i}} \cdot \frac{\partial E}{\partial \epsilon_{j}} \bigg\rangle
    - \bigg\langle \frac{\partial E}{\partial \epsilon_i} \bigg\rangle \bigg\langle \frac{\partial E}{\partial \epsilon_j} \bigg\rangle \biggr\} \\
    &= \sum^{N}_{r=1} \biggl\{ \frac{\partial^{2} E(X_{r} | \epsilon)}{\partial \epsilon_{i}\partial \epsilon_{j}}
        - \bigg\langle \frac{\partial^{2} E}{\partial \epsilon_{i}\partial \epsilon_{j}} \bigg\rangle + \operatorname{Cov} \!\left( \frac{\partial E}{\partial \epsilon_i}, \frac{\partial E}{\partial \epsilon_j} \right) \biggr\}.
  \end{split}
\end{equation}
And when $i=j$, the second-order partial derivative of the energy function becomes
\begin{equation}
\begin{split}
    \frac{\partial^{2} u}{\partial \epsilon^{2}} = \sum^{N}_{r=1} \biggl\{& \frac{\partial^{2} E(X_{r}|\epsilon)}{\partial \epsilon^{2}} - \bigg\langle \frac{\partial^{2} E(X|\epsilon)}{\partial \epsilon^{2}} \bigg\rangle \\
    &+ \bigg\langle \left(\frac{\partial E(X|\epsilon)}{\partial \epsilon}\right)^{2} \bigg\rangle  - \bigg\langle \frac{\partial E(X|\epsilon)}{\partial \epsilon} \bigg\rangle^{2}
    \biggr\}.
\end{split}
\label{eq:a16}
\end{equation}


\section{Kofinger \& Hummer's toy polymer model}
\label{sec:polymer_model}
As mentioned in the main text, the two dimensional polymer model\cite{kofinger2021empirical} consists of a string of $n=100$ beads separated by a unit of length, where $\Delta\phi_{i}$ is the angle between the two bond vectors $v_{i}$ and $v_{i+1}$. The associated probability of a polymer with $n$ beads is given by
\begin{equation}
  \begin{split}
    p(\boldsymbol{\Delta} \boldsymbol{\phi} \mid \boldsymbol{\mu}, \boldsymbol{\kappa}) &=\prod_{i=1}^{n-1} p\left(\Delta \phi_{i} \mid \mu_{i}, \kappa_{i}\right) \\
    &=\prod_{i=1}^{n-1}\frac{\exp( \kappa_{i} \cos \left(\Delta \phi_{i}-\mu_{i}\right))}{2 \pi I_{0}\left(\kappa_{i}\right)}
  \end{split}
\end{equation}
where $I_{0}(\kappa)$ is the modified Bessel function of order zero and $\mu_{i}$ introduces directional biases.  The parameters $\kappa_{i} > 0$ determine the stiffness of the polymer at positions $i$.  Prior energy, $E$ of the polymer configuration follow the Boltzmann distribution, which corresponds to
\begin{equation}
E = - \log \prod_{i=1}^{n-1} p\left(\Delta \phi_{i} \mid \mu_{i}, \kappa_{i}\right)
\end{equation}
Next, we briefly derive derivatives of the toy model energy function and prior energies that will be used in our algorithm.  The first derivative of the prior energy $E$ with respect to $\kappa$ is
\begin{equation}
  \begin{split}
    \frac{\partial E}{\partial \kappa} &= - \sum_{i=1}^{n-1} \frac{ \partial }{\partial \kappa} \log p\left(\Delta \phi_{i} \mid \mu_{i}, \kappa_{i}\right) \\
    &= - \sum_{i=1}^{n-1} \frac{ 1 }{p\left(\Delta \phi_{i} \mid \mu_{i}, \kappa_{i}\right)} \frac{ \partial p\left(\Delta \phi_{i} \mid \mu_{i}, \kappa_{i}\right)}{\partial \kappa},
  \end{split}
  \label{eq:d_VM_prior}
\end{equation}
where the derivative of the von Mises distribution with respect to $\kappa$ is
\begin{equation}
  \begin{split}
    \frac{\partial p}{\partial \kappa} = &\frac{ \exp\left(\kappa_{i} \cos(\mu_{i} - \Delta \phi_{i})\right) \cos(\mu_{i} - \Delta \phi_{i}) }{2 \pi I_{0}\left(\kappa_{i}\right)}  \\
    &\quad- \frac{ \exp\left(\kappa_{i} \cos(\mu_{i} - \Delta \phi_{i})\right) I_{1}\left(\kappa_{i}\right) }{2 \pi {I_{0}}^{2}\left(\kappa_{i}\right)}
  \end{split}
\end{equation}
which is relatively straightforward since the derivative of the modified Bessel function merely increases it's order.
Next, the second derivative of $E$ with respect to $\kappa_{i}$:
\begin{equation}
  \frac{\partial^2 E}{\partial \kappa_{i}^2} = \sum \left\{-\frac{1}{p} \frac{\partial^2 p}{\partial \kappa_{i}^2}+\left(\frac{1}{p} \frac{\partial p}{\partial \kappa_{i}}\right)^2 \right\},
  \label{eq:d2_VM_prior}
\end{equation}
where the second derivative of the von Mises distribution w.r.t $\kappa_{i}$ is
\begin{equation}
  \begin{aligned}
    \frac{\partial^{2} p}{\partial \kappa_{i}^{2}} = \biggl[& \frac{ \exp\left(\kappa_{i} \cos(\mu_{i} - \Delta \phi_{i})\right) \left(\frac{I_{0}\left(\kappa_{i}\right)}{2} + \frac{I_{2}\left(\kappa_{i}\right)}{2} \right)}{2 \pi {I_{0}}^{2}\left(\kappa_{i}\right)}  \\
    &+ \frac{ \exp\left(\kappa_{i} \cos(\mu_{i} - \Delta \phi_{i})\right) \cos^{2}(\mu_{i} - \Delta \phi_{i}) }{2 \pi I_{0}\left(\kappa_{i}\right)} \\
    & - \frac{ \exp\left(\kappa_{i} \cos(\mu_{i} - \Delta \phi_{i})\right) \cos(\mu_{i} - \Delta \phi_{i})I_{1}\left(\kappa_{i}\right) }{ \pi {I_{0}}^{2}\left(\kappa_{i}\right)}\\
    & + \frac{ \exp\left(\kappa_{i} \cos(\mu_{i} - \Delta \phi_{i})\right) {I_{1}}^{2}\left(\kappa_{i}\right) }{ \pi {I_{0}}^{3}\left(\kappa_{i}\right)}\biggr]
  \end{aligned}
\end{equation}

\bibliography{references}

\section*{TOC Graphic}

\begin{figure}[htb!]
\centering
  \includegraphics[width=0.75\linewidth]{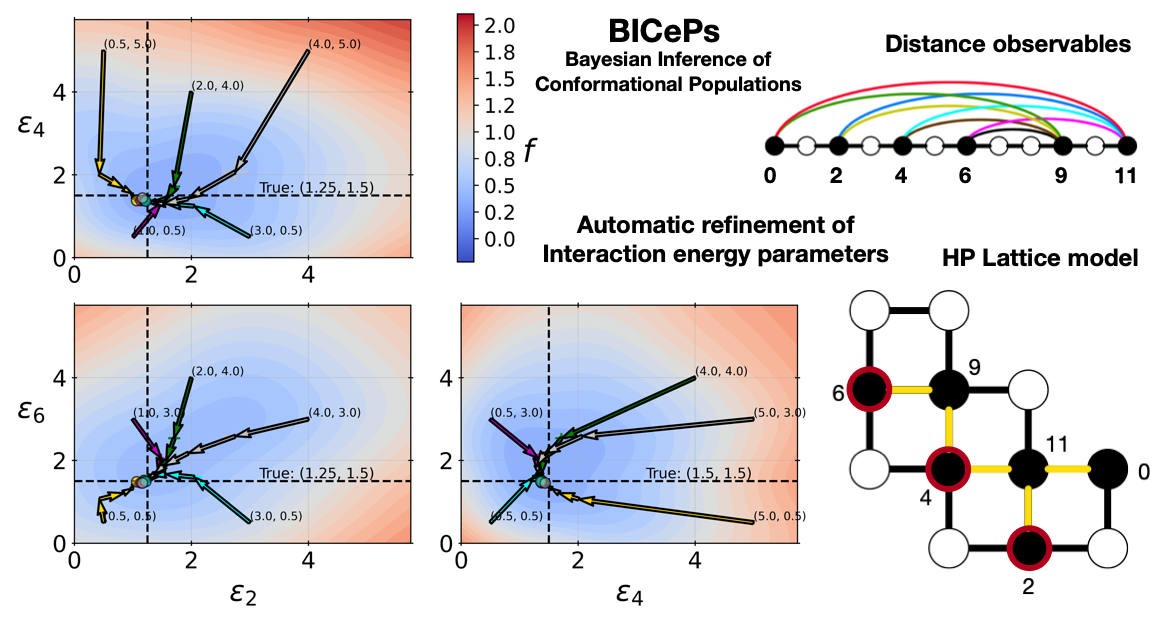}
\end{figure}

\newpage

\include{SI-aip.tex}

\end{document}

%% file: SI-aip.tex
\newpage
\clearpage

\onecolumngrid
\setcounter{page}{1}
\renewcommand{\thepage}{S\arabic{page}}

\setcounter{figure}{0}
\renewcommand{\thefigure}{S\arabic{figure}}

\setcounter{table}{0}
\renewcommand{\thetable}{S\arabic{table}}

\setcounter{equation}{0}
\renewcommand{\theequation}{S\arabic{equation}}

\begin{center}
{\Large\bfseries Supplemental Material}

\vspace{1cm}

{\large\bf Automated optimization of force field parameters against ensemble-averaged measurements with Bayesian Inference of Conformational Populations}

\vspace{0.5cm}

Robert M. Raddi and Vincent A. Voelz

\vspace{0.25cm}

Department of Chemistry, Temple University,
Philadelphia, Pennsylvania 19122, USA

\end{center}

\vspace{1cm}

This PDF contains additional figures, tables, and computational details
supporting the results presented in the main text.

\bigskip
\newpage


\begin{figure*}
\centering
  \includegraphics[width=\linewidth]{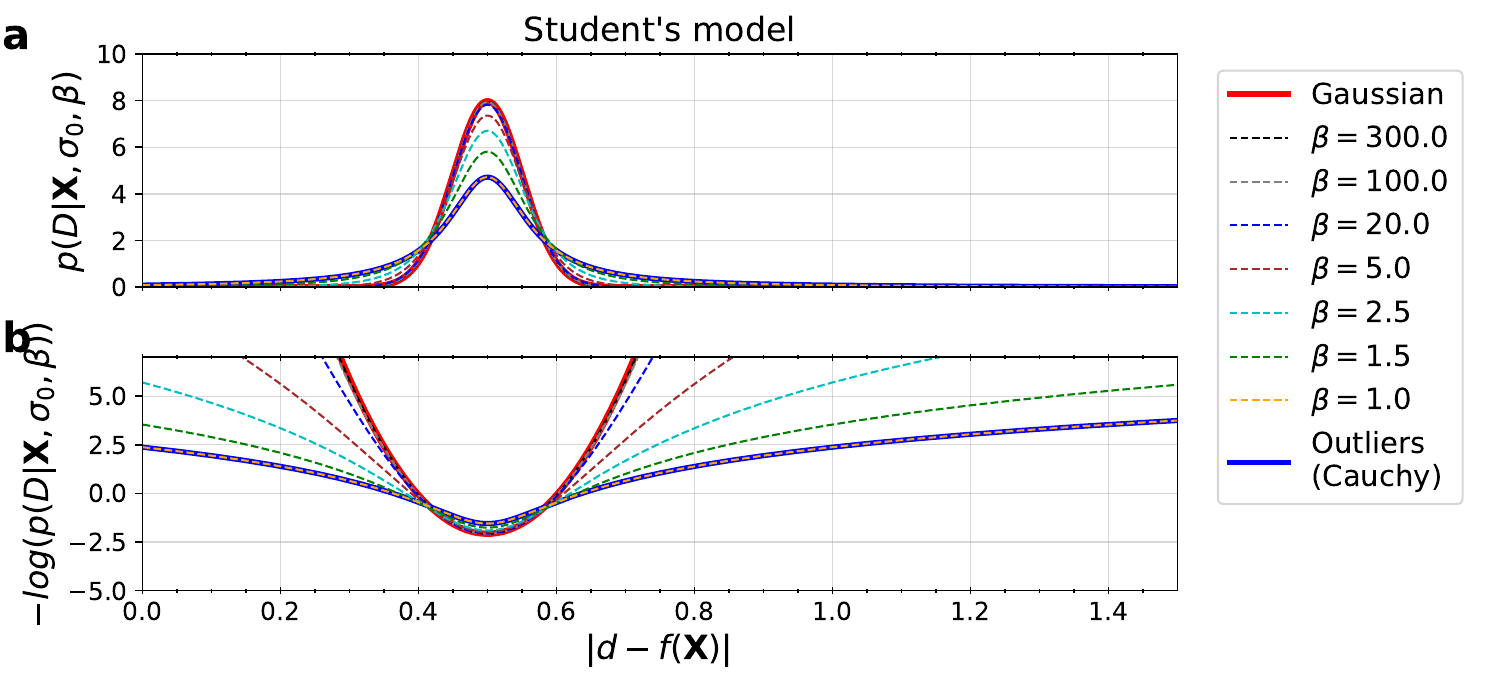}
  \caption{\small The probability density function (a) and energy landscape (b) of the marginal likelihood for the Student's model with respect to the replica-averaged forward model data $f(\mathbf{X})$. The colored curves are different values of nuisance parameter $\beta$. The Student's model is equivalent to the Outliers model (Cauchy) when $\beta=1$, and as $\beta$ goes to infinity, the pdf becomes Gaussian and the energy becomes harmonic.}
  \label{fig:Students_restraint_marginal_likelihood}
\end{figure*}


\begin{figure*}
\centering
  \includegraphics[width=\linewidth]{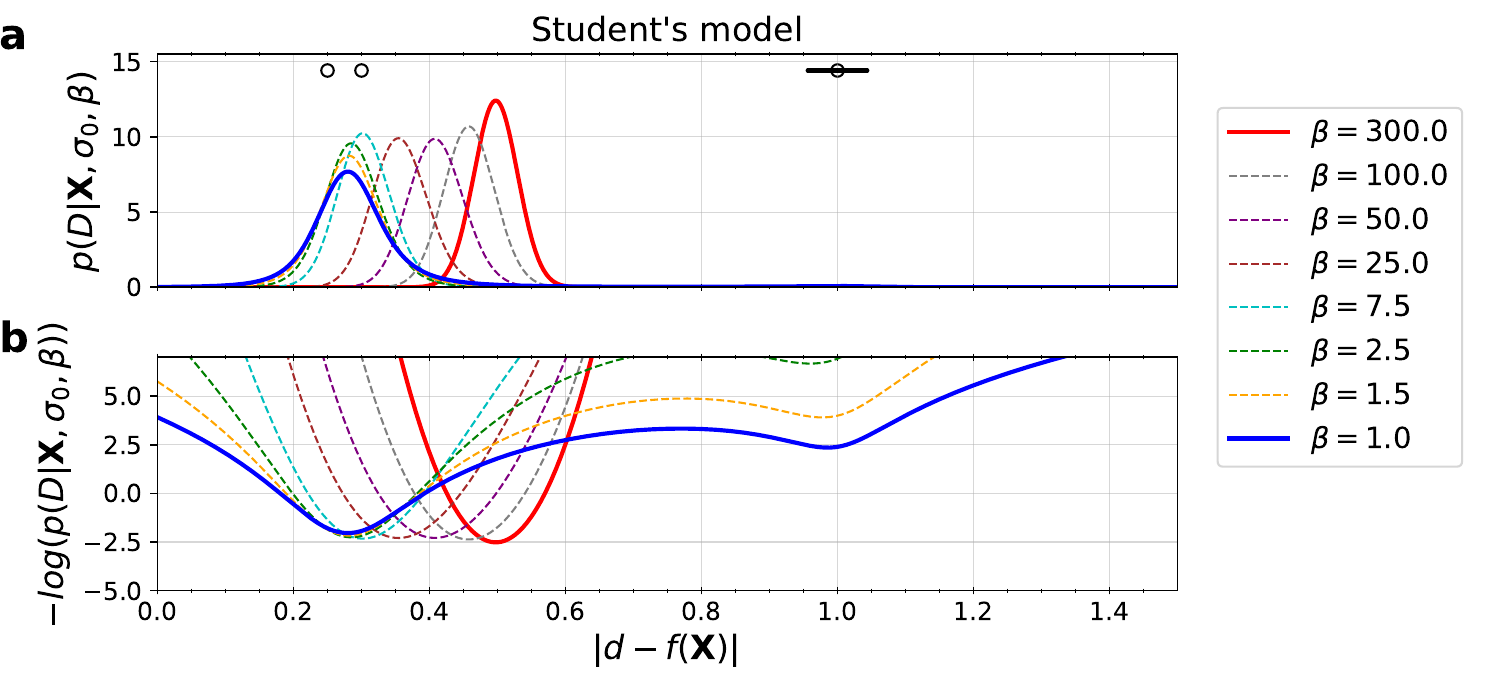}
  \caption{\small The probability density function (a) computed as $p(D|\mathbf{X}, \sigma_{0},\beta) = \prod^{N_{j}}_{j} p(d_{j}|\mathbf{X}, \sigma_{0},\beta)$ and energy landscape (b) of the marginal likelihood for the Student's model with respect to the replica-averaged forward model data $f(\mathbf{X})$ using multiple data points. Shown here, are three data points, two good data points \{0.25, 0.3\} and one outlier \{1.0\}. The Cauchy likelihood ($\beta=1$) is centered about the mean of the two good data points, demonstrating that this model can distinguish the good and bad data. The standard Gaussian likelihood ($\beta=300$) is centered about the mean of all three data points. The colored curves are different values of nuisance parameter $\beta$. }
  \label{fig:Students_example}
\end{figure*}

\begin{figure*}
\centering
  \includegraphics[width=\linewidth]{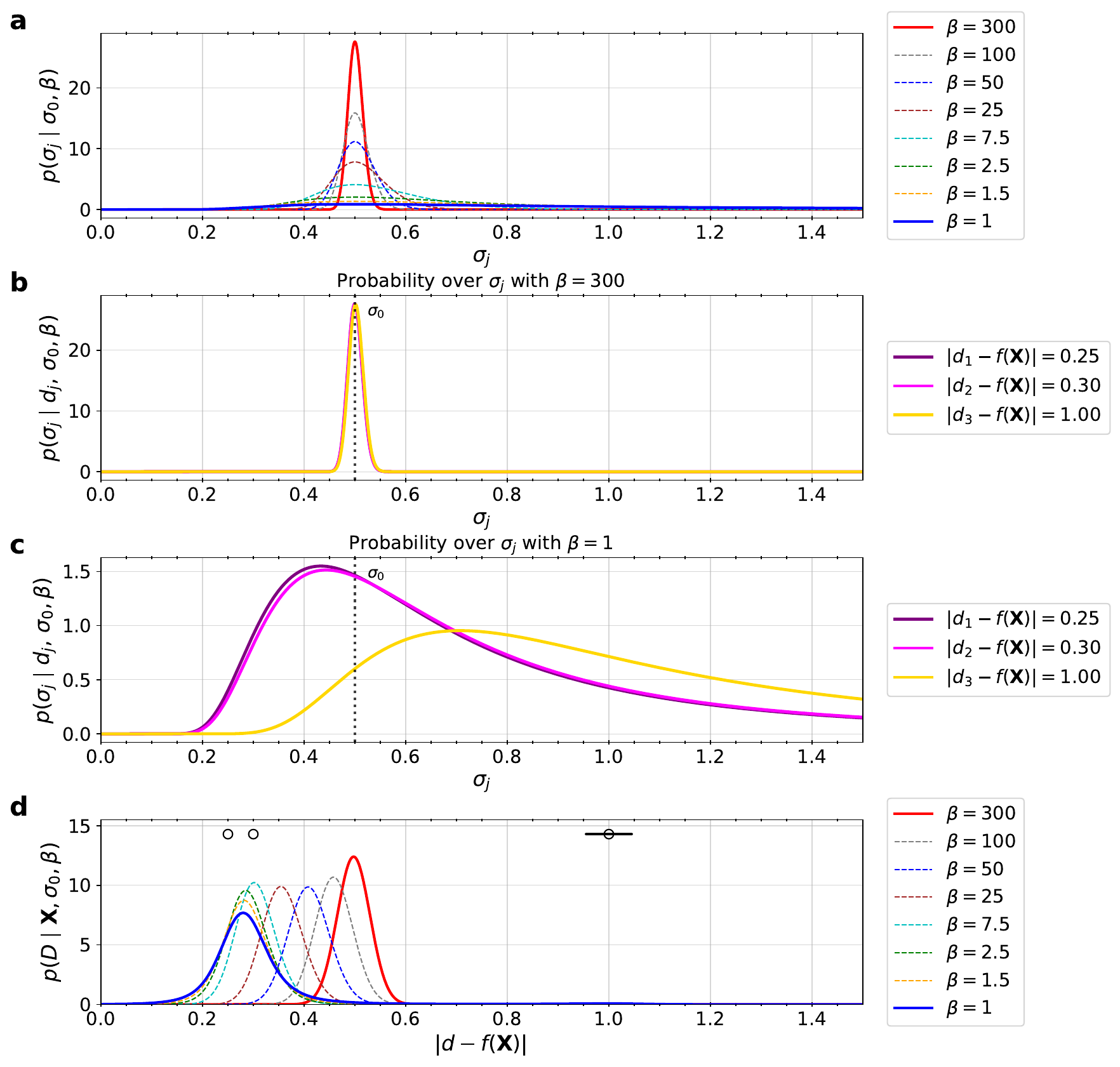}
  \caption{\small Hierarchical interpretation of the Student’s likelihood model across the prior on uncertainties, latent uncertainty, and the marginalized likelihood.  (a) Prior distribution over the latent uncertainties, $p(\sigma_j \mid \sigma_0,\beta)$, for a range of $\beta$ values. The parameter $\beta$ controls the tail behavior of the distribution: small $\beta$ yields broader, heavy-tailed priors, while large $\beta$ concentrates probability near $\sigma_j \approx \sigma_0$, approaching the Gaussian limit as $\beta \to \infty$. (b-c) Posterior distribution over $\sigma_j$ for individual data points, $p(\sigma_j \mid d_j,\sigma_0,\beta) \propto p(d_j \mid \sigma_j)\,p(\sigma_j \mid \sigma_0,\beta)$, shown for two typical observations and one outlier using $\beta=300$ and $\beta=1$, respectively. The good data points ($|d_1 - f(\mathbf{X})| =$ 0.25 and $|d_2 - f(\mathbf{X})| =$ 0.30) favor values of $\sigma_j$ near $\sigma_0$, whereas the outlier ($|d_3 - f(\mathbf{X})| =$ 1.0) shifts probability toward larger $\sigma_j$, effectively inflating the local uncertainty. (d) Marginalized likelihood, $p(D \mid \mathbf{X},\sigma_0,\beta) = \prod_j \int p(d_j \mid \sigma_j)\,p(\sigma_j \mid \sigma_0,\beta)\,d\sigma_j,$ plotted as a function of the residual $|d - f(\mathbf{X})|$. The open circles mark the observed data (two good data points and one outlier). Decreasing $\beta$ produces heavier tails in the likelihood, allowing the model to accommodate outliers without strongly penalizing deviations, while increasing $\beta$ recovers a sharply peaked Gaussian-like likelihood.
 }\label{fig:Students_prior_example}
\end{figure*}

\begin{figure*}
\centering
  \includegraphics[width=\linewidth]{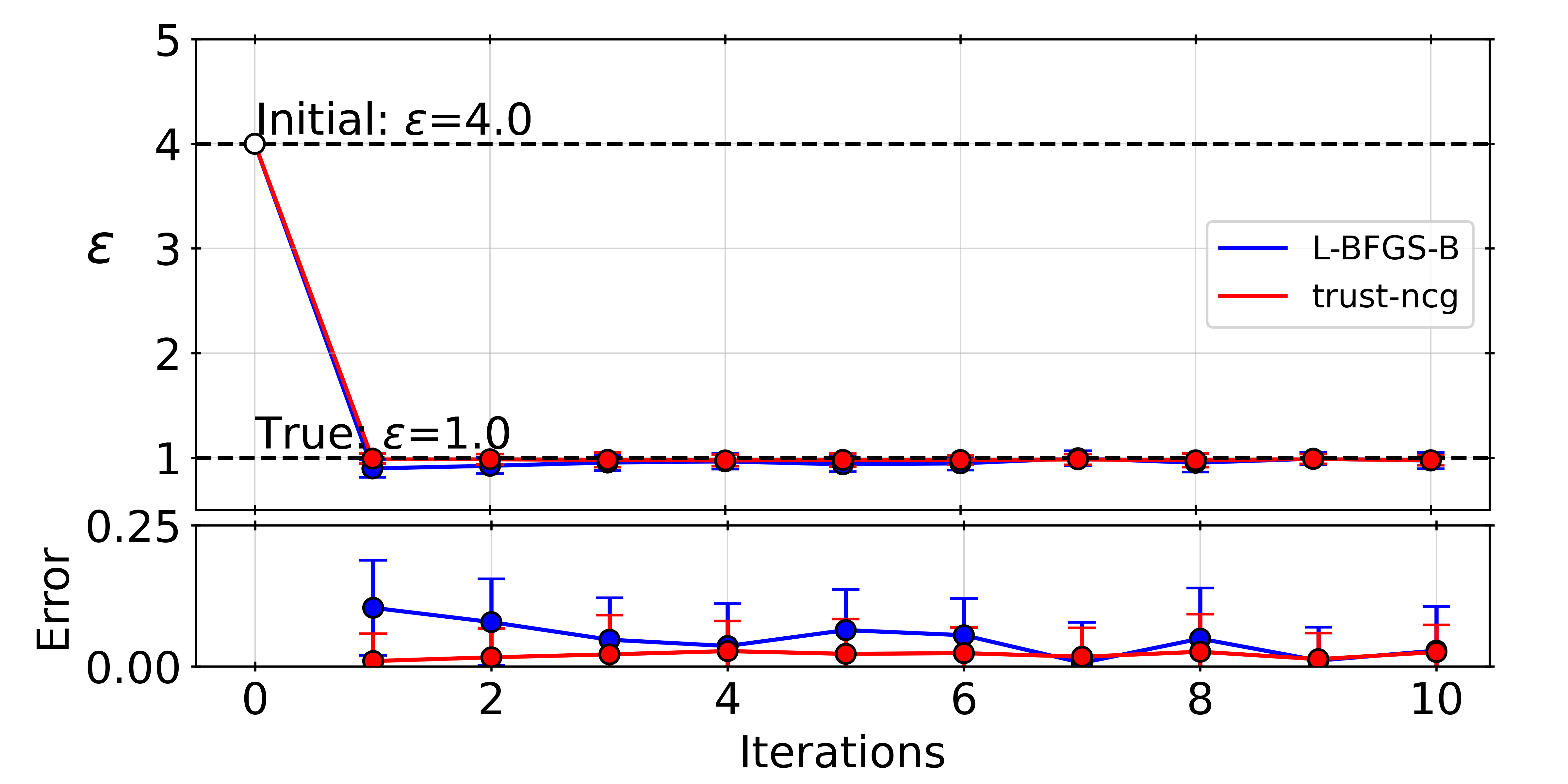}
  \caption{\small Trust-region optimization with BICePs is found to converge faster and be more robust than L-BFGS-B. Here, we compare first-order L-BFGS-B method (blue) to the second-order Trust-NCG optimization method (red). A total of 25 independent rounds of parameter ($\epsilon$) optimizations using the BICePs score for a maximum of ten iterations. Data points are placed at the mean value for each iteration and error bars are the standard deviation. It is typical that the trust-ncg method reaches convergence slightly faster (the 4th iteration), and the L-BFGS-B method tends to slightly overshoot the "True" $\epsilon$ value on the first iteration.}
  \label{fig:bootstrapped_1-D_optimization}
\end{figure*}

\begin{figure*}
  \centering
  \includegraphics[width=\linewidth]{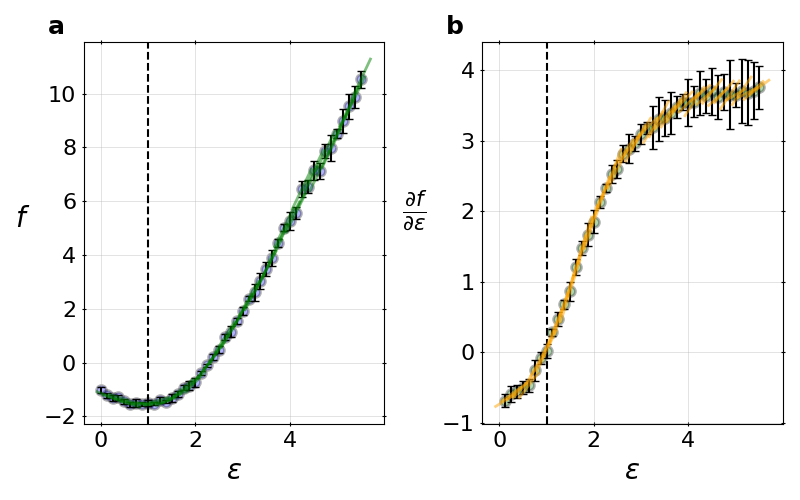}
  \caption{\small
      (a) 1-D scans over $\epsilon$ to reveal the landscape of the BICePs score (scatter dots). The green tangent lines are the derivatives at each $\epsilon$ value. Uncertainties in the BICePs scores and derivatives come from the standard error of the mean across five independent scans along $\epsilon$. (b) The derivative of the BICePs score (scatter dots) and the second derivative of the BICePs score (orange tangent lines) at each epsilon value. The dotted black line at $\epsilon^{*}=1.0$ shows the true value, which is where the derivative of the BICePs score equals zero. BICePs calculations are run using the Student's likelihood model for 100k steps with 8 replicas.
    }
  \label{fig:score_landscape}
\end{figure*}

\begin{figure*}
\centering
  \includegraphics[width=\linewidth]{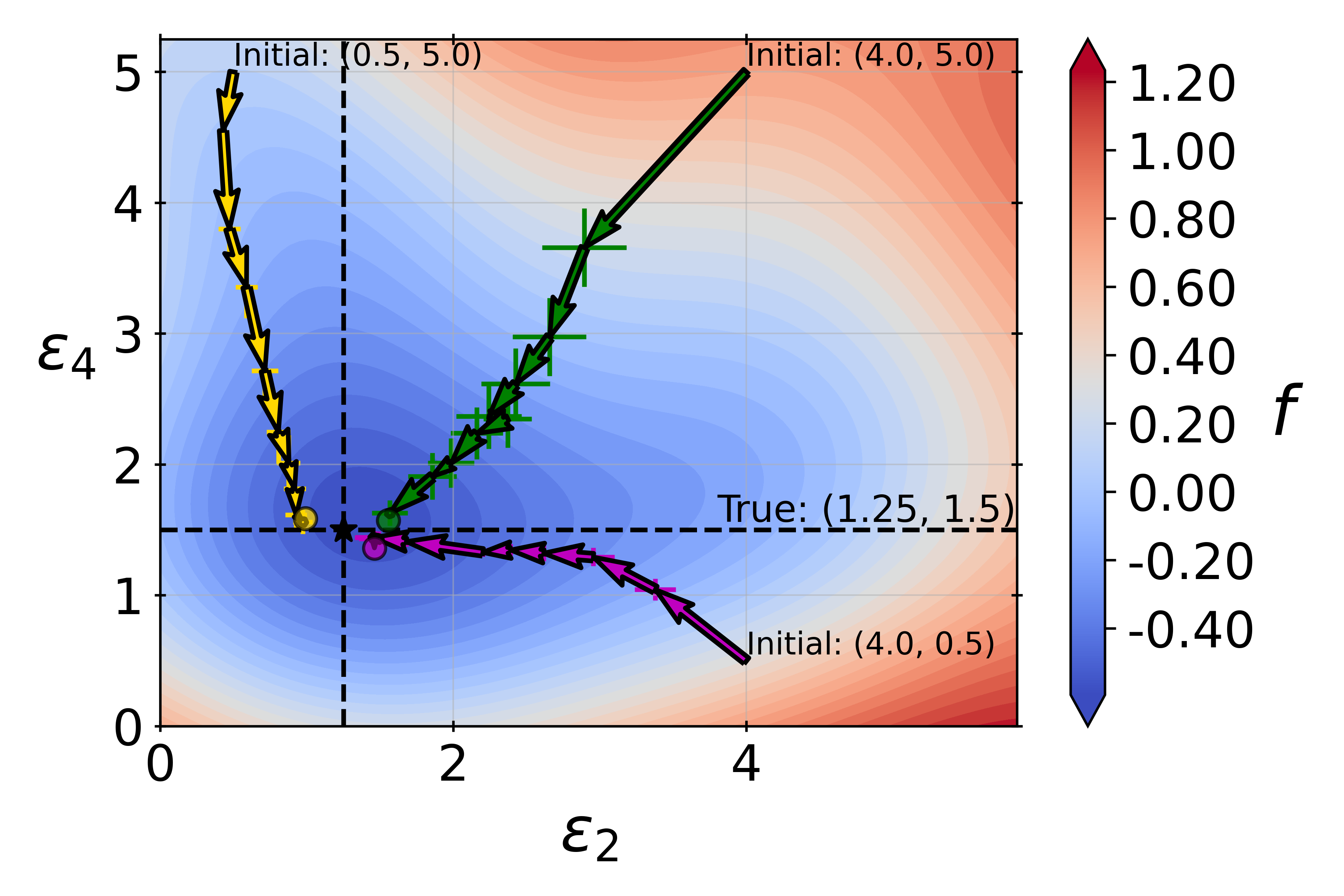}
  \caption{\small Average traces over a total of 25 independent rounds of parameter $(\epsilon_{2}, \epsilon_{4})$ optimizations using first-order optimization method (L-BFGS-B) with BICePs, for a maximum of ten iterations. Traces are unable to fully reach the "True" iteration strength parameters ($\epsilon_{2}^{*}=1.25$, $\epsilon_{4}^{*}=1.5$) within 10 iterations when starting from different initial parameters $(\epsilon_{2}^{0}, \epsilon_{4}^{0})$ = $\{(0.5, 5.0), (4.0, 5.0), (4.0, 0.5)\}$.  The BICePs score landscape was generated from the average values of five scans over $\epsilon_{2}$ and $\epsilon_{4}$. All calculations used the Student's model with 200k MCMC steps and 8 replicas.
  }
  \label{fig:2-D_L-BFGS-B}
\end{figure*}

\begin{figure*}
\centering
  \includegraphics[width=\linewidth]{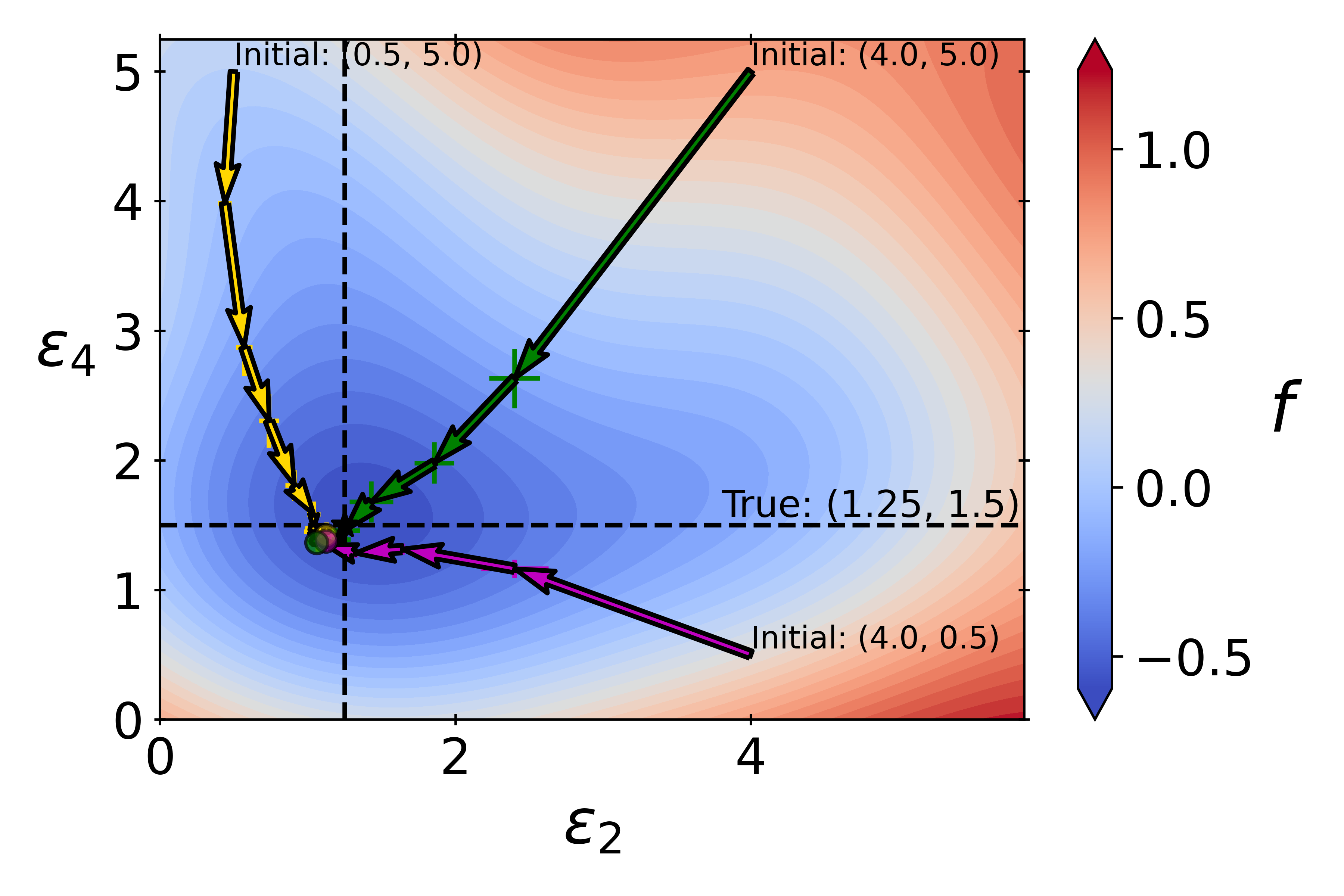}
  \caption{\small Average traces over a total of 25 independent rounds of parameter $(\epsilon_{2}, \epsilon_{4})$ optimizations using second-order (trust-ncg) method with BICePs, for a maximum of ten iterations. Optimizations converge to the "True" optimal iteration strength parameters ($\epsilon_{2}^{*}=1.25$, $\epsilon_{4}^{*}=1.5$) when starting from different initial parameters $(\epsilon_{2}^{0}, \epsilon_{4}^{0})$ = $\{(0.5, 5.0), (4.0, 5.0), (4.0, 0.5)\}$.  The BICePs score landscape was generated from the average values of five scans over $\epsilon_{2}$ and $\epsilon_{4}$. All calculations used the Student's model with 200k MCMC steps and 8 replicas.  Average optimized parameter values were determined to be ($\epsilon_2$ = 1.07 $\pm$ 0.85, and $\epsilon_4$ = 1.46 $\pm$ 0.82), where the uncertainties are estimated from the inverse Hessian.
  }
  \label{fig:2-D_trust-ncg}
\end{figure*}

\begin{figure*}
\centering
  \includegraphics[width=\linewidth]{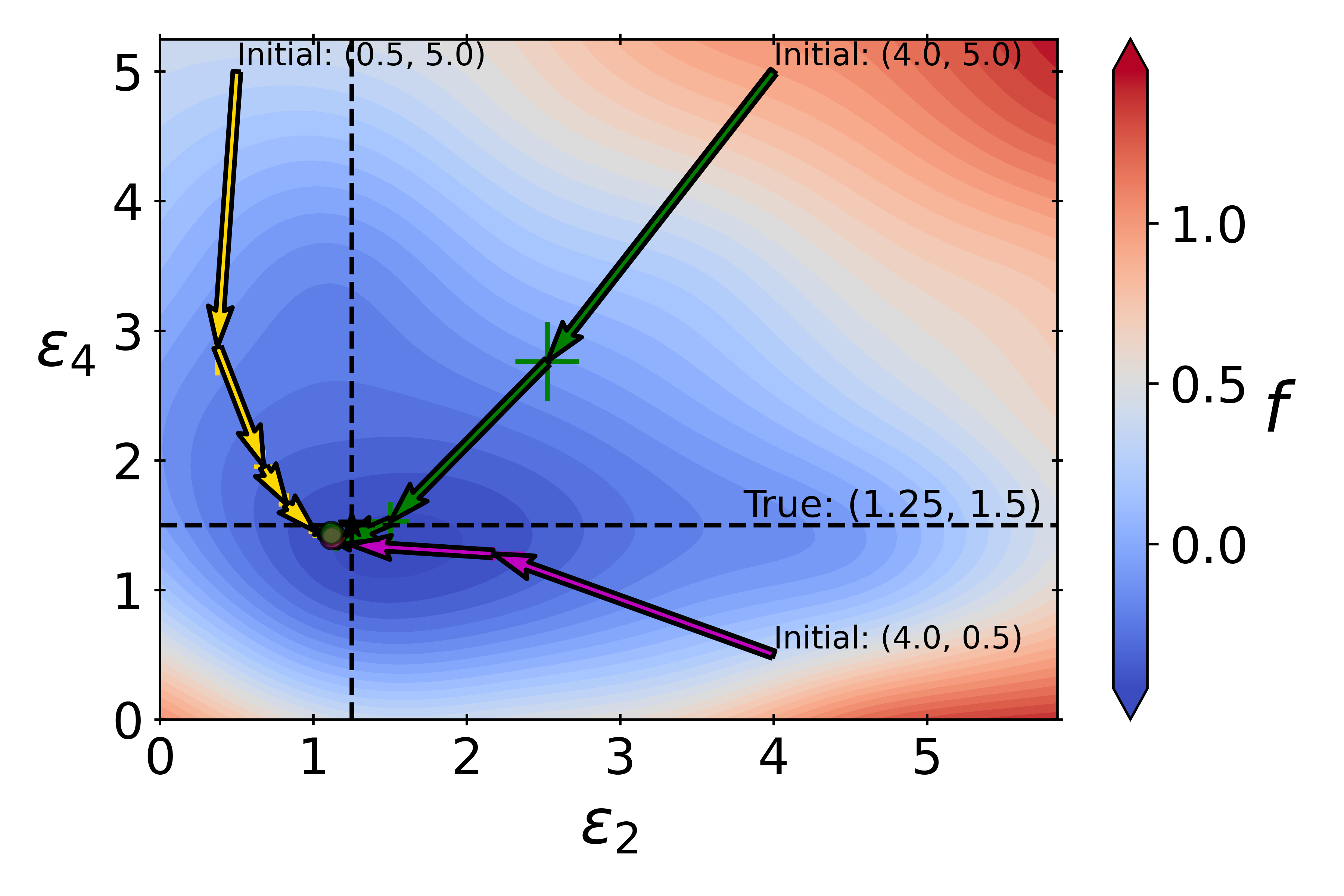}
  \caption{\small Average traces over a total of 25 independent rounds of parameter $(\epsilon_{2}, \epsilon_{4})$ optimizations using second-order (trust-ncg) method with BICePs, for a maximum of ten iterations. In these tests, no error was added to the experimental data. Optimizations converge to the "True" parameters ($\epsilon_{2}^{*}=1.25$, $\epsilon_{4}^{*}=1.5$) when starting from different initial parameters $(\epsilon_{2}^{0}, \epsilon_{4}^{0})$ = $\{(0.5, 5.0), (4.0, 5.0), (4.0, 0.5)\}$.  The BICePs score landscape was generated from the average values of five scans over $\epsilon_{2}$ and $\epsilon_{4}$. All calculations used the Student's model with 200k MCMC steps and 32 replicas. Average optimized parameter values were
    ($\epsilon_2$ = 1.12 $\pm$ 0.42, and $\epsilon_4$ = 1.43 $\pm$ 0.36),
  where the uncertainties are estimated from the inverse Hessian.
  }
  \label{fig:2-D_opt_epsilon_2_and_4_no_error}
\end{figure*}

\begin{figure*}
\centering
  \includegraphics[width=\linewidth]{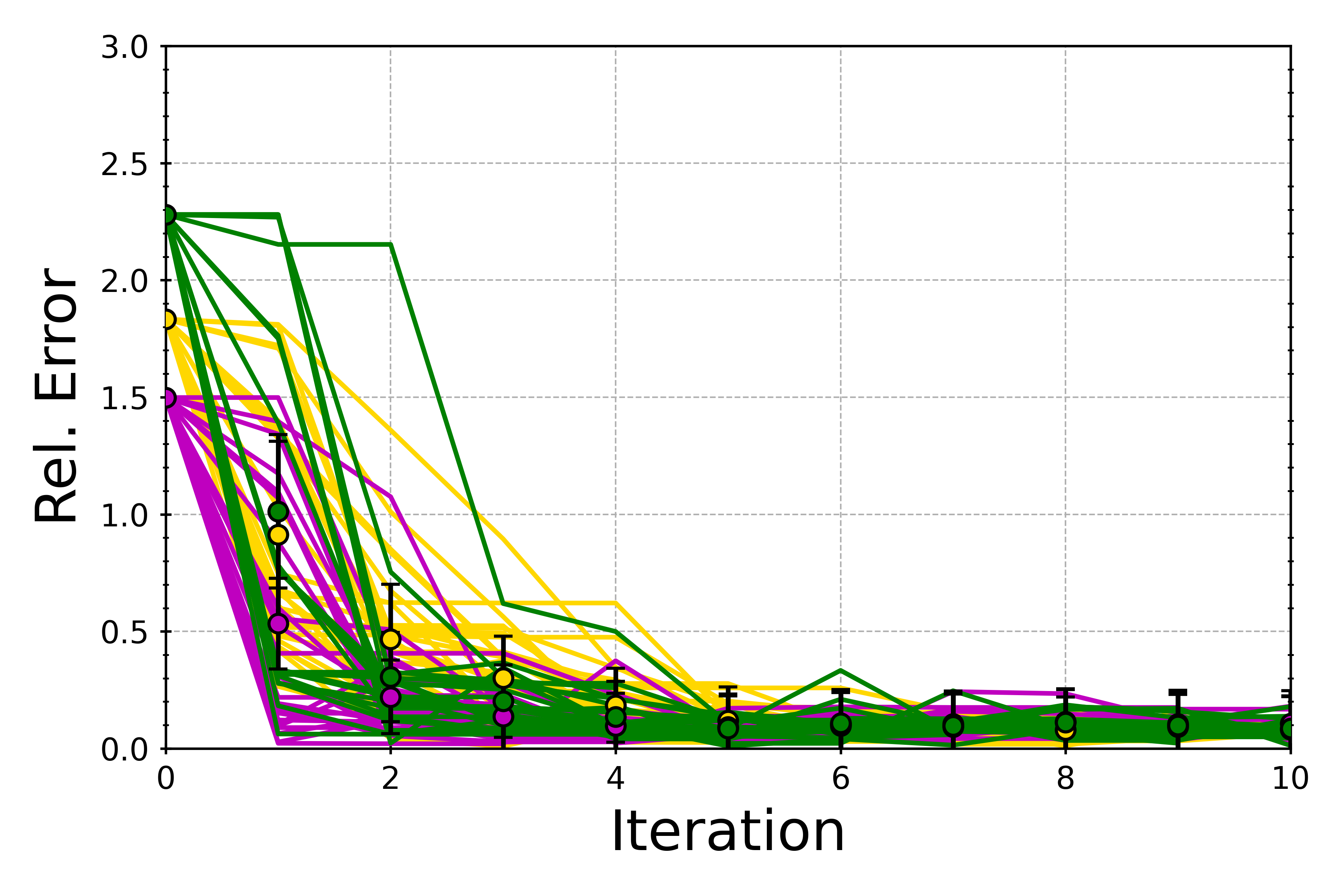}
  \caption{\small Accuracy profiles for $(\epsilon_{2}, \epsilon_{4})$ optimizations using second-order (trust-ncg) method with BICePs, for a total of 25 independent rounds and a maximum of ten iterations. In these tests, no error was added to the experimental data.
  Optimizations converge to the "True" parameters ($\epsilon_{2}^{*}=1.25$, $\epsilon_{4}^{*}=1.5$) with a relative error less than 1.0 when starting from different initial parameters $(\epsilon_{2}^{0}, \epsilon_{4}^{0})$ = $\{(0.5, 5.0), (4.0, 5.0), (4.0, 0.5)\}$.  All calculations used the Student's model with 200k MCMC steps and 32 replicas.  Average optimized parameter values were
    ($\epsilon_2$ = 1.12 $\pm$ 0.42, and $\epsilon_4$ = 1.43 $\pm$ 0.36),
  where the uncertainties are estimated from inverse Hessian. Error bars represent uncertainties from Monte Carlo error propagation using the $\sigma_{\epsilon}$ from the inverse Hessian, taken over all independent optimizations.
  }
  \label{fig:accuracy_profile_32_rep_epsilon_2_and_4_no_error}
\end{figure*}

\begin{figure*}
\centering
  \includegraphics[width=\linewidth]{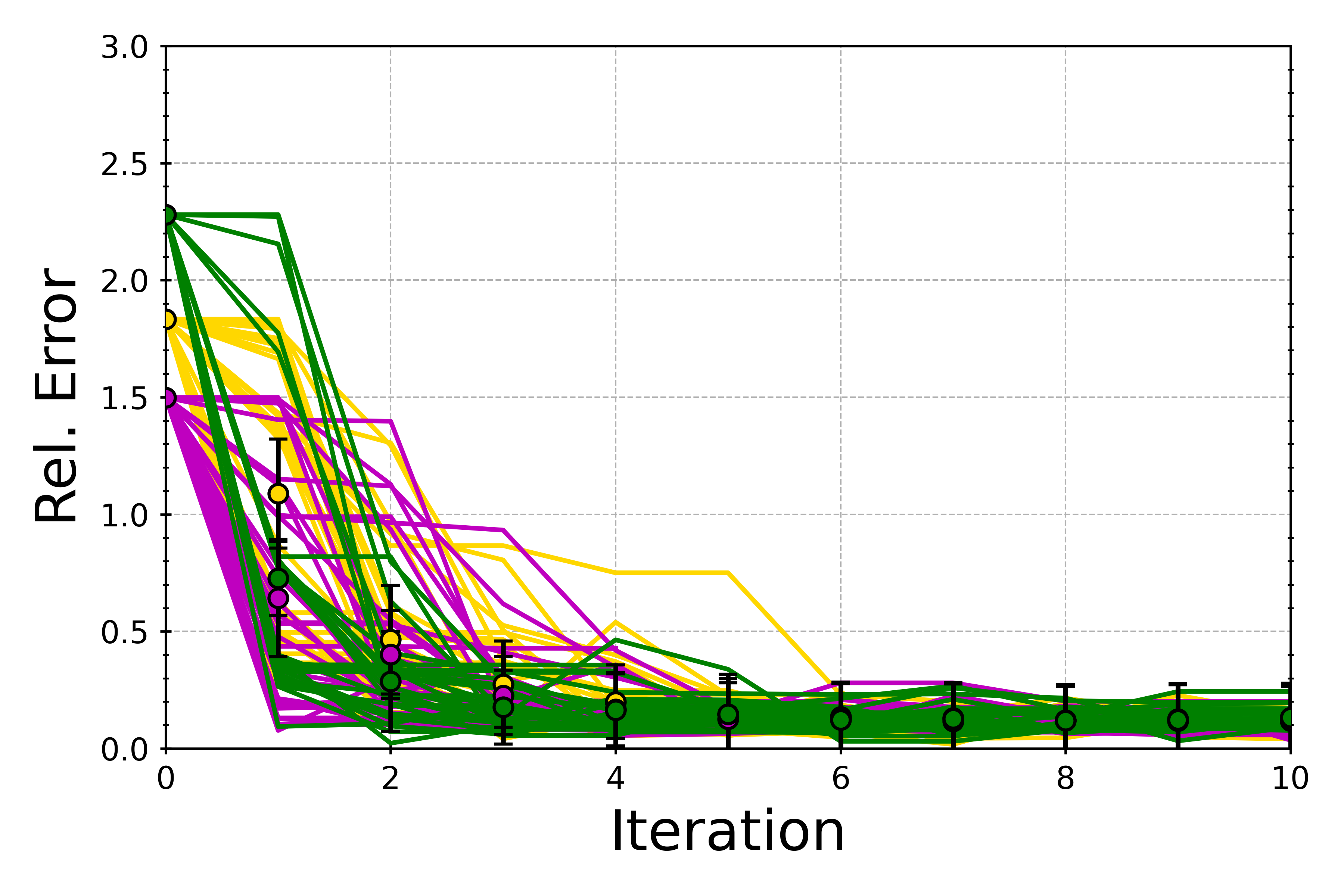}
  \caption{\small Accuracy profiles for $(\epsilon_{2}, \epsilon_{4})$ optimizations using second-order (trust-ncg) method with BICePs, for a total of 25 independent rounds and a maximum of ten iterations. In these tests, the experimental data was corrupted with systematic error in the 2--11 and 4--9 distances for +3 and +3.5 L.U. shift, respectively.  The total error in the data is $\sigma_{data} = 1.63$ L.U..   Optimizations converge to the "True" optimal iteration strength parameters ($\epsilon_{2}^{*}=1.25$, $\epsilon_{4}^{*}=1.5$) with a relative error less than 1.0 when starting from different initial parameters $(\epsilon_{2}^{0}, \epsilon_{4}^{0})$ = $\{(0.5, 5.0), (4.0, 5.0), (4.0, 0.5)\}$. Error bars represent  uncertainties estimated from the inverse Hessian, averaged over the 25 independent optimizations.
  All calculations used the Student's model with 200k MCMC steps and 32 replicas.
  }
  \label{fig:accuracy_profile_32_rep_epsilon_2_and_4_systematic_error}
\end{figure*}

\begin{figure*}
\centering
  \includegraphics[width=\linewidth]{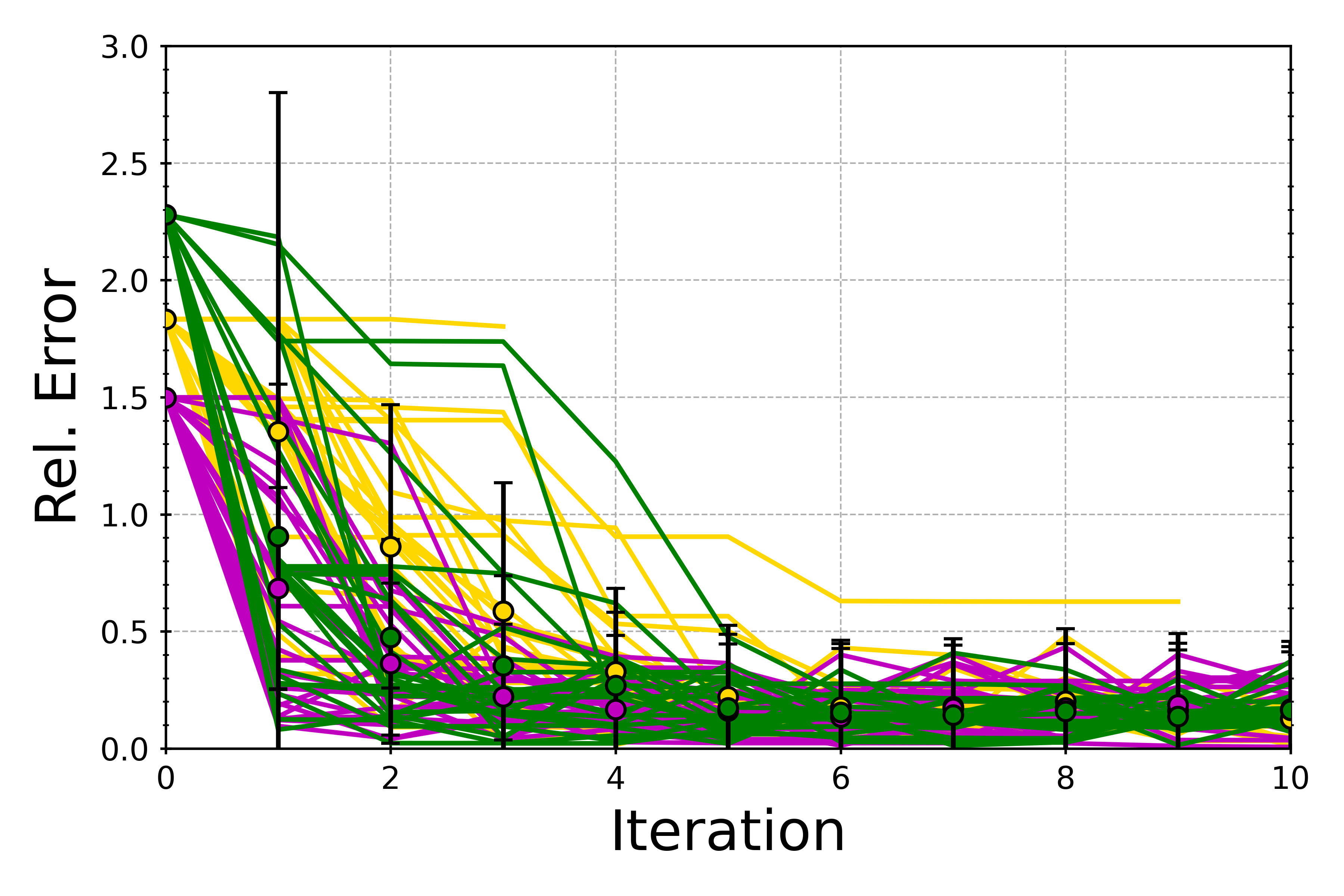}
  \caption{\small Accuracy profiles for $(\epsilon_{2}, \epsilon_{4})$ optimizations using a second-order (trust-ncg) method with BICePs, for a total of 25 independent rounds and a maximum of ten iterations. In these tests, no error was added to the experimental data.
  Optimizations converge to the "True" parameters ($\epsilon_{2}^{*}=1.25$, $\epsilon_{4}^{*}=1.5$) with a relative error less than 1.0 when starting from different initial parameters $(\epsilon_{2}^{0}, \epsilon_{4}^{0})$ = $\{(0.5, 5.0), (4.0, 5.0), (4.0, 0.5)\}$.  Error bars represent  uncertainties estimated from the inverse Hessian, averaged over the 25 independent optimizations.  All calculations used the Student's model with 200k MCMC steps and 8 replicas. Average optimized parameter values were determined to be  ($\epsilon_2$ = 1.07 $\pm$ 0.85, and $\epsilon_4$ = 1.46 $\pm$ 0.82), where the uncertainties are estimated from inverse Hessian. Error bars represent uncertainties from Monte Carlo error propagation using the $\sigma_{\epsilon}$ from the inverse Hessian, taken over all independent optimizations.
  }
  \label{fig:accuracy_profile_8_rep_epsilon_2_and_4}
\end{figure*}

\begin{figure*}
\centering
  \includegraphics[width=0.9\linewidth]{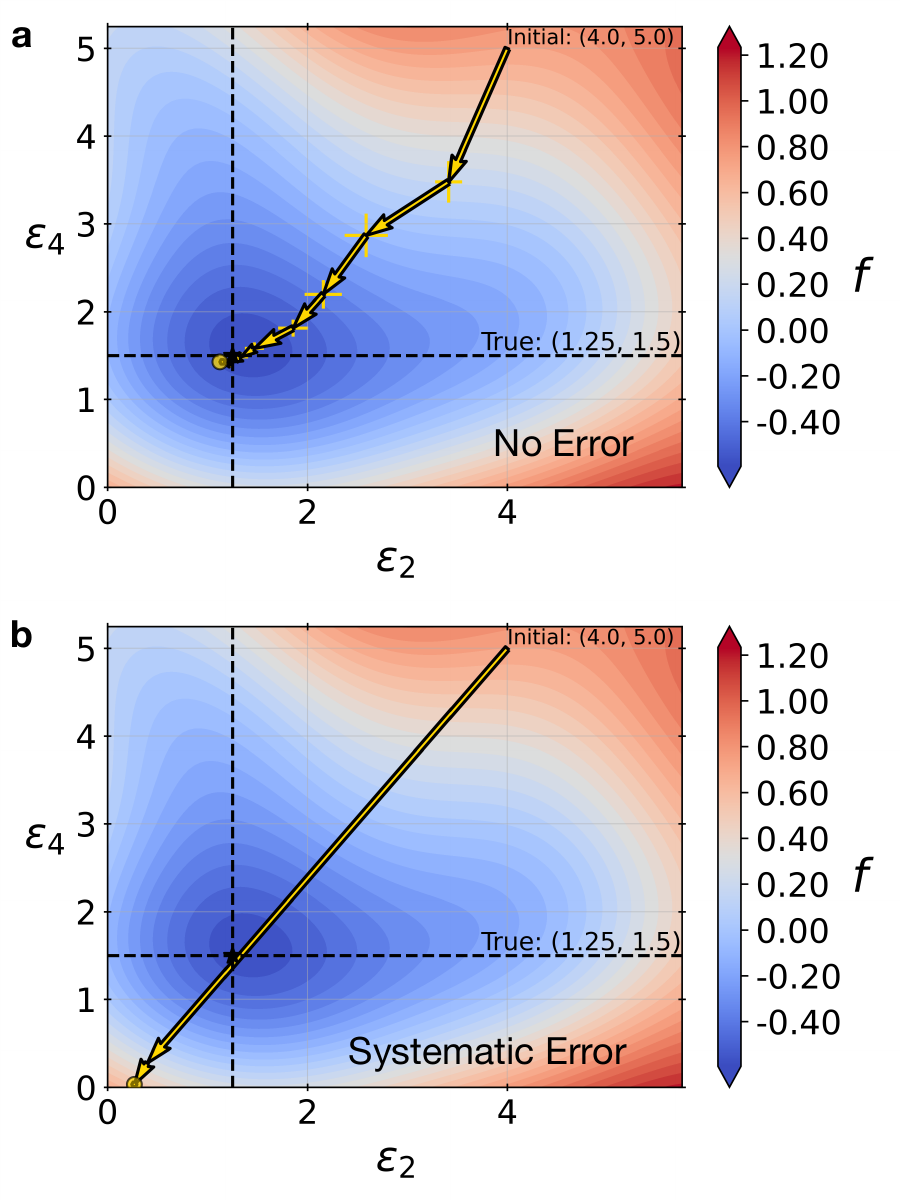}
  \caption{\small  The accuracy of a standard Gaussian likelihood drops in the presence of systematic error. Demonstrating optimization accuracy when the data is corrupted with and without systematic error using $\epsilon_{2}=4.0$ and $\epsilon_{4}=5.0$ as initial parameters, same conditions as in Figure \ref{fig:2-D_opt_epsilon_2_and_4}, where the "True" parameters are $(\epsilon_{2}^{*}=1.25, \epsilon_{4}^{*}=1.5)$. (a) The Gaussian likelihood performs great when using experimental data not affected by systematic error. (b) The accuracy suffers when the data is corrupted with systematic error in the 2-11 and 4-9 distances for +3 and +3.5 L.U. shift, respectively.  The total error in the data is $\sigma_{data} = 1.63$. All results shown here use 8 replicas and are averages over 25 independent optimizations, and error bars are computed as standard error of the mean for each iteration.}
  \label{fig:2-D_landscape_systematic_error_Gaussian}
\end{figure*}

\begin{figure*}
\centering
  \includegraphics[width=0.9\linewidth]{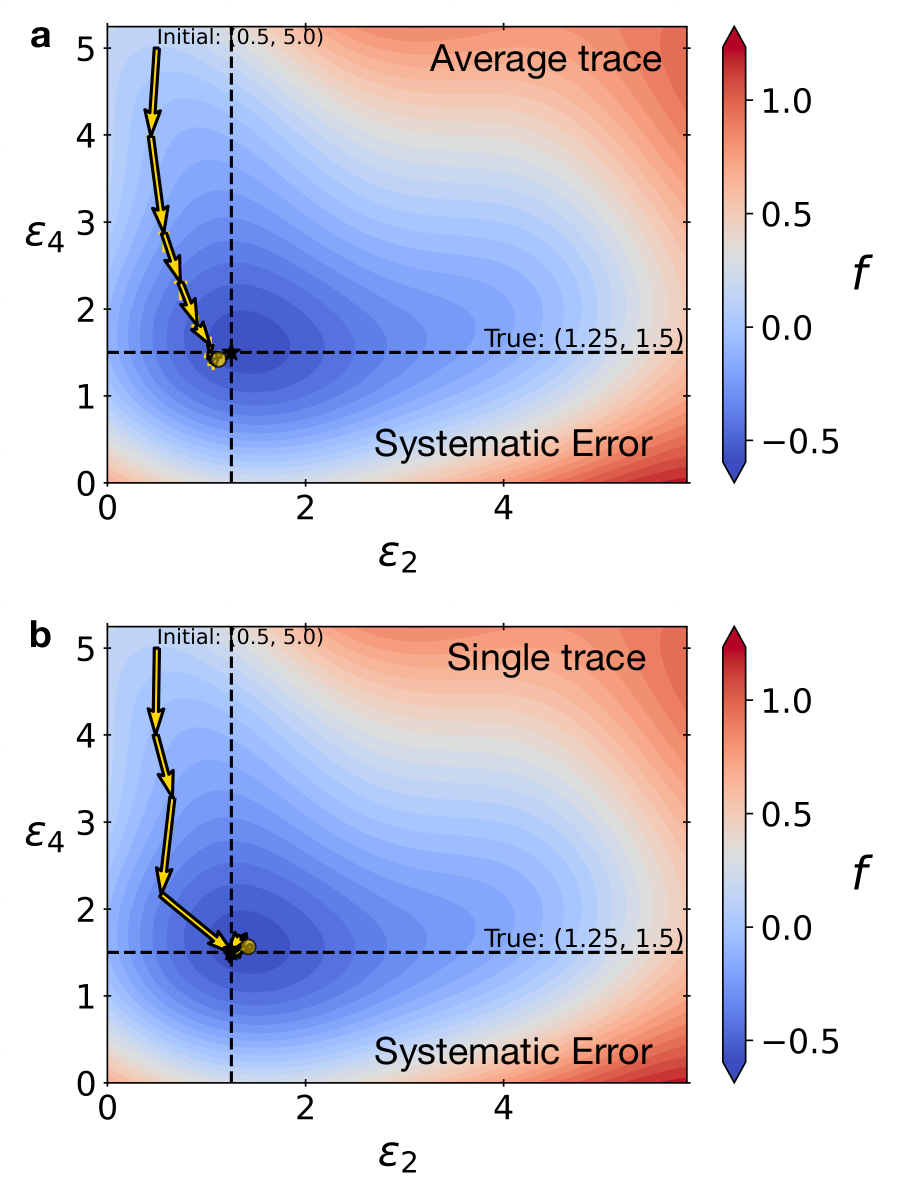}
  \caption{\small  The accuracy of the Student's model is very robust in the presence of systematic error. (a) Average traces over 25 independent optimizations when the data is corrupted with systematic error in the 2--11 and 4--9 distances for +3 and +3.5 L.U. shift, respectively. The total error in the data is $\sigma_{data} = 1.63$.  (b) A single representative trace demonstrating convergence to the "True" parameters $(\epsilon_{2}^{*}=1.25, \epsilon_{4}^{*}=1.5)$. All results shown here use 8 replicas and are averages over 25 independent optimizations.
  }
  \label{fig:2-D_landscape_systematic_error_Students}
\end{figure*}

\begin{figure*}
\centering
  \includegraphics[width=\linewidth]{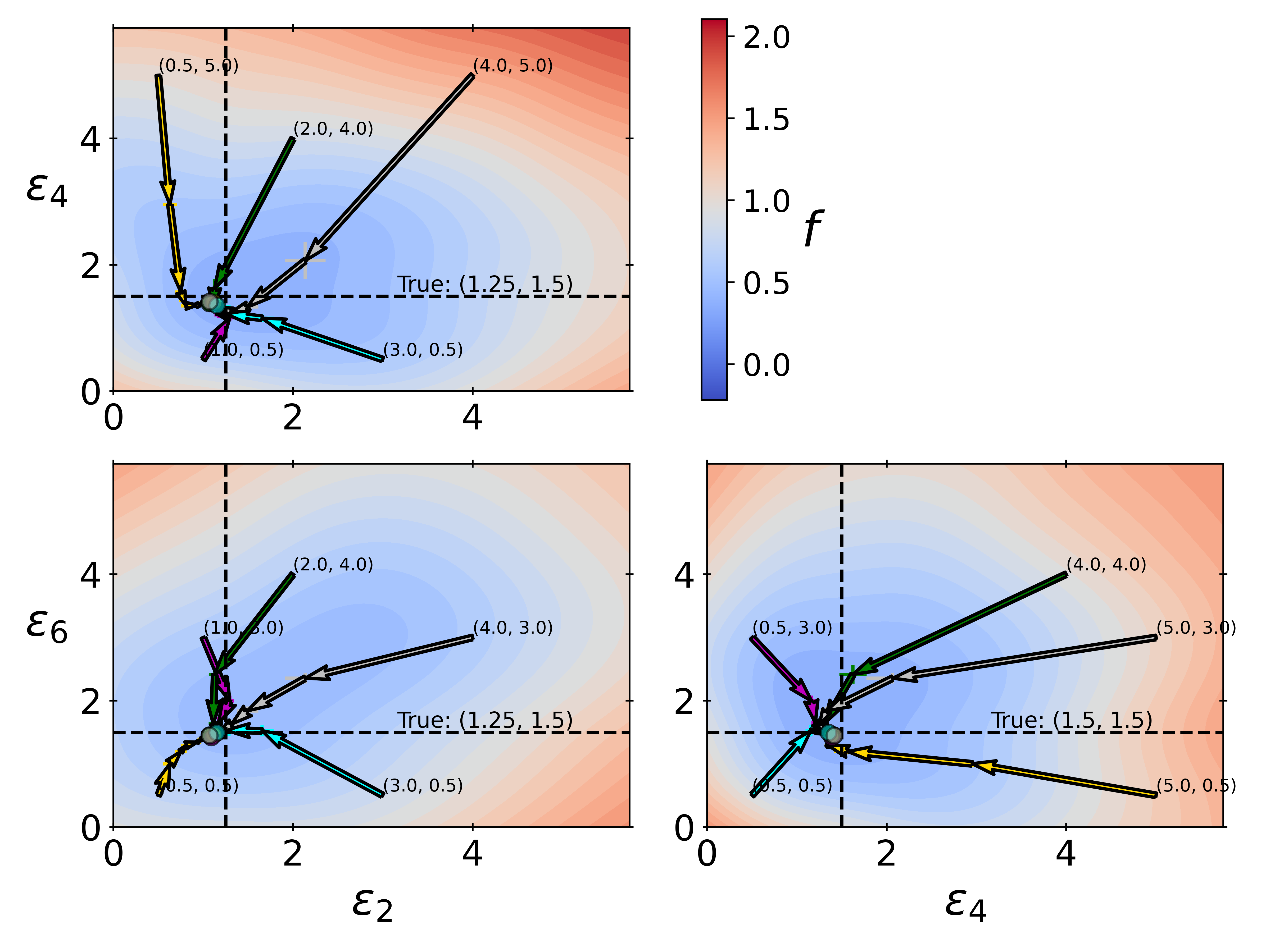}
  \caption{\small  Average traces over a total of 25 independent rounds of parameter $(\epsilon_{2}, \epsilon_{4}, \epsilon_{6})$ optimizations using second-order (trust-ncg) method with BICePs, for a maximum of ten iterations. Optimizations converge to the "True" optimal iteration strength parameters ($\epsilon_{2}^{*}=1.25$, $\epsilon_{4}^{*}=1.5$, $\epsilon_{6}^{*}=1.5$) when starting from different initial parameters. All calculations used the Student's model with 200k MCMC steps and 32 replicas.
  }
  \label{fig:3-D_opt__epsilon_2_4_6}
\end{figure*}

\begin{figure*}
\centering
  \includegraphics[width=0.9\linewidth]{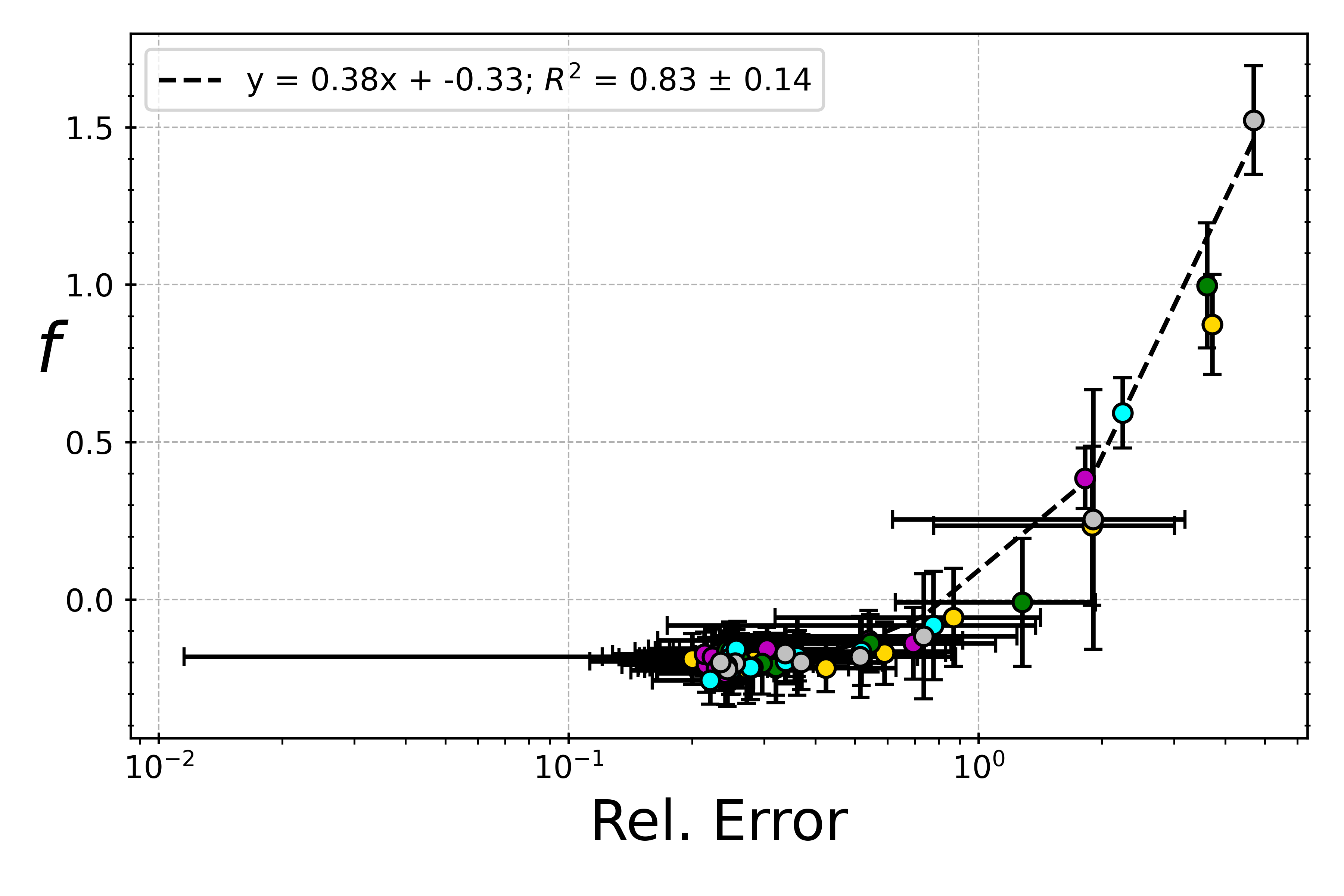}
  \caption{\small  Sensitivity analysis of parameter $(\epsilon_{2}, \epsilon_{4}, \epsilon_{6})$ optimizations from various initial conditions. Regardless of different starting parameters, optimizations universally converge to "True" optimal iteration strength parameters $(\epsilon_{2}^{*}=1.25$, $\epsilon_{4}^{*}=1.5$, $\epsilon_{6}^{*}=1.5$) and succeed in diminishing the relative error. All BICePs score optimizations used 32 BICePs replicas with the Student's likelilhood model, and 200k MCMC steps per iteration. The dashed line is a linear fitting with equation $y = 0.38x -0.33$, resulting in a strong correlation between the BICePs score and relative model error (coefficient of determination $R^{2} = 0.83 \pm 0.14$).
  }
  \label{fig:3-D_sensitivity}
\end{figure*}

\begin{figure*}
\centering
  \includegraphics[width=0.9\linewidth]{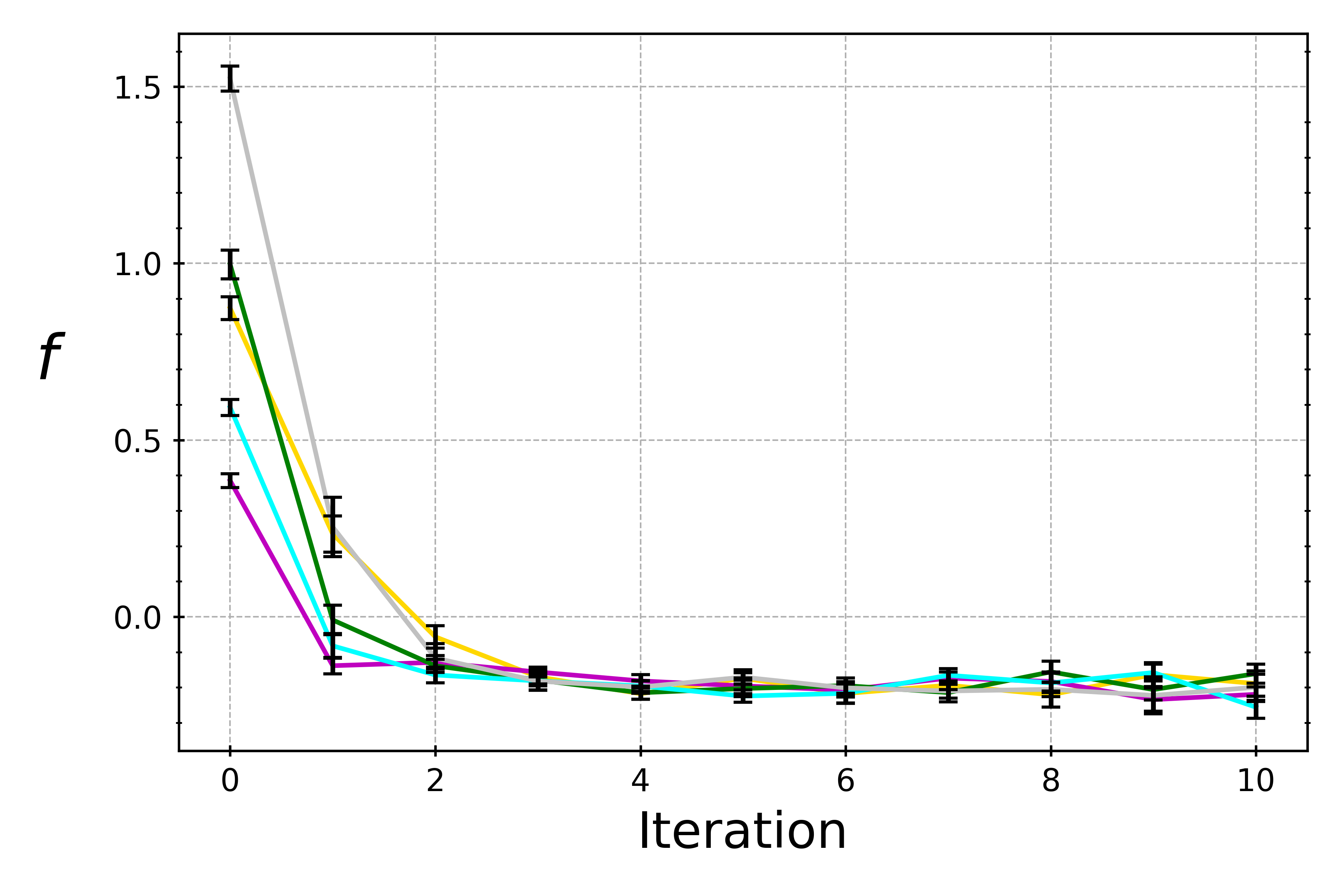}
  \caption{\small  Minimization of the BICePs score, $f$ is demonstrated when averaging over 25 independent rounds of parameter $(\epsilon_{2}, \epsilon_{4}, \epsilon_{6})$ optimizations using second-order (trust-ncg) method, for a maximum of ten iterations. Error bars represent the standard error over the 25 optimizations.
  }
  \label{fig:3-D_score_vs_iteration}
\end{figure*}

\begin{figure*}
\centering
  \includegraphics[width=0.9\linewidth]{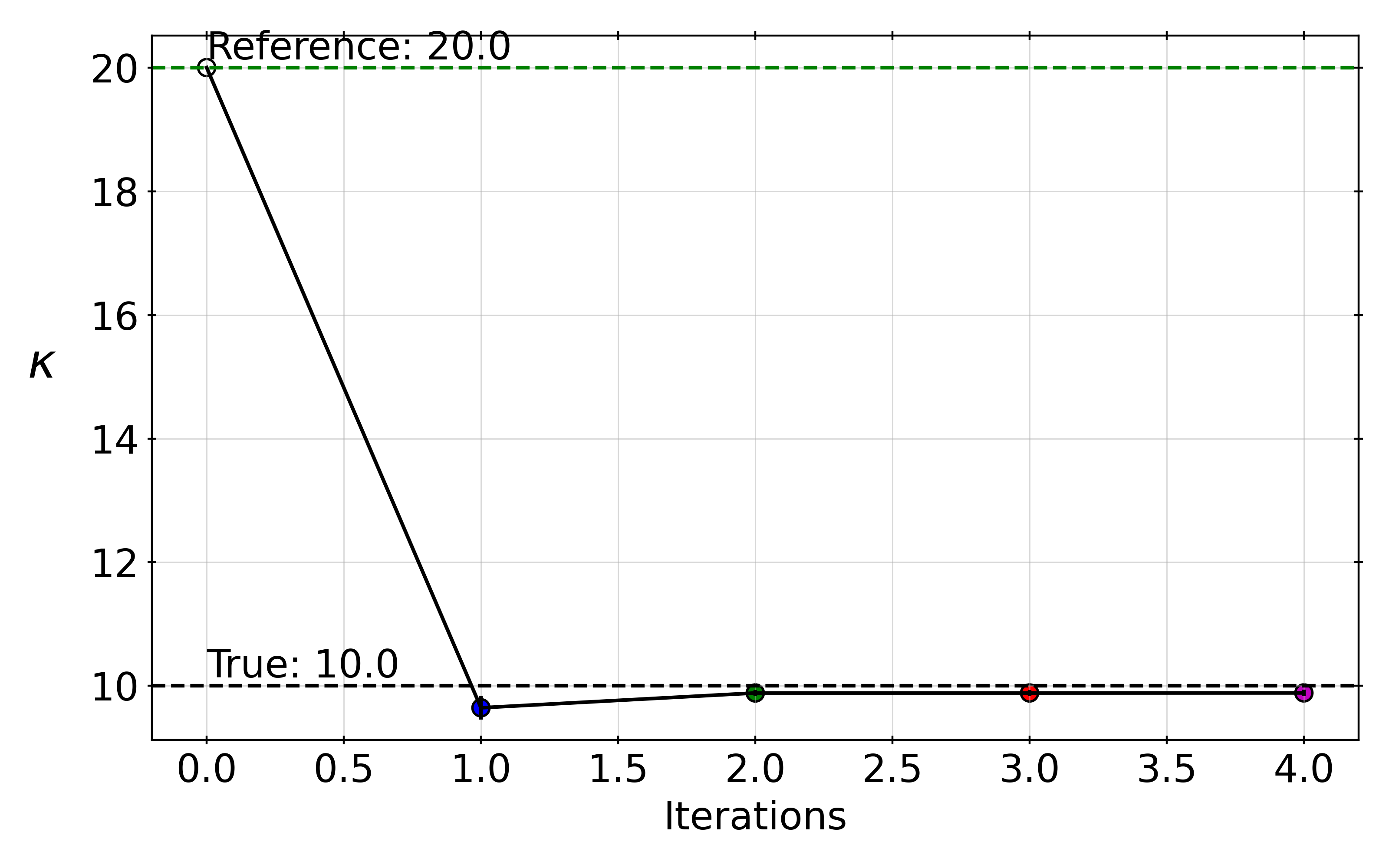}
  \caption{\small  Refinement of the $\kappa$, the stiffness parameter in the KH polymer model. An initial parameter value, $\kappa^{0}=20$, was established. Employing second-order optimization via the `trust-ncg` method, convergence to the “true” parameter, $\kappa^{*}=10$, was achieved within a few iterations. The computations used 200 replicas and 100k MCMC steps, encompassing 500 configurations. Parameter uncertainties throughout each iteration were estimated from the inverse Hessian. These results successfully replicate the outcome of BioFF by Kofinger and Hummer.\cite{kofinger2021empirical}.
  }
  \label{fig:polymer_model}
\end{figure*}

\begin{figure*}
\centering
  \includegraphics[width=0.9\linewidth]{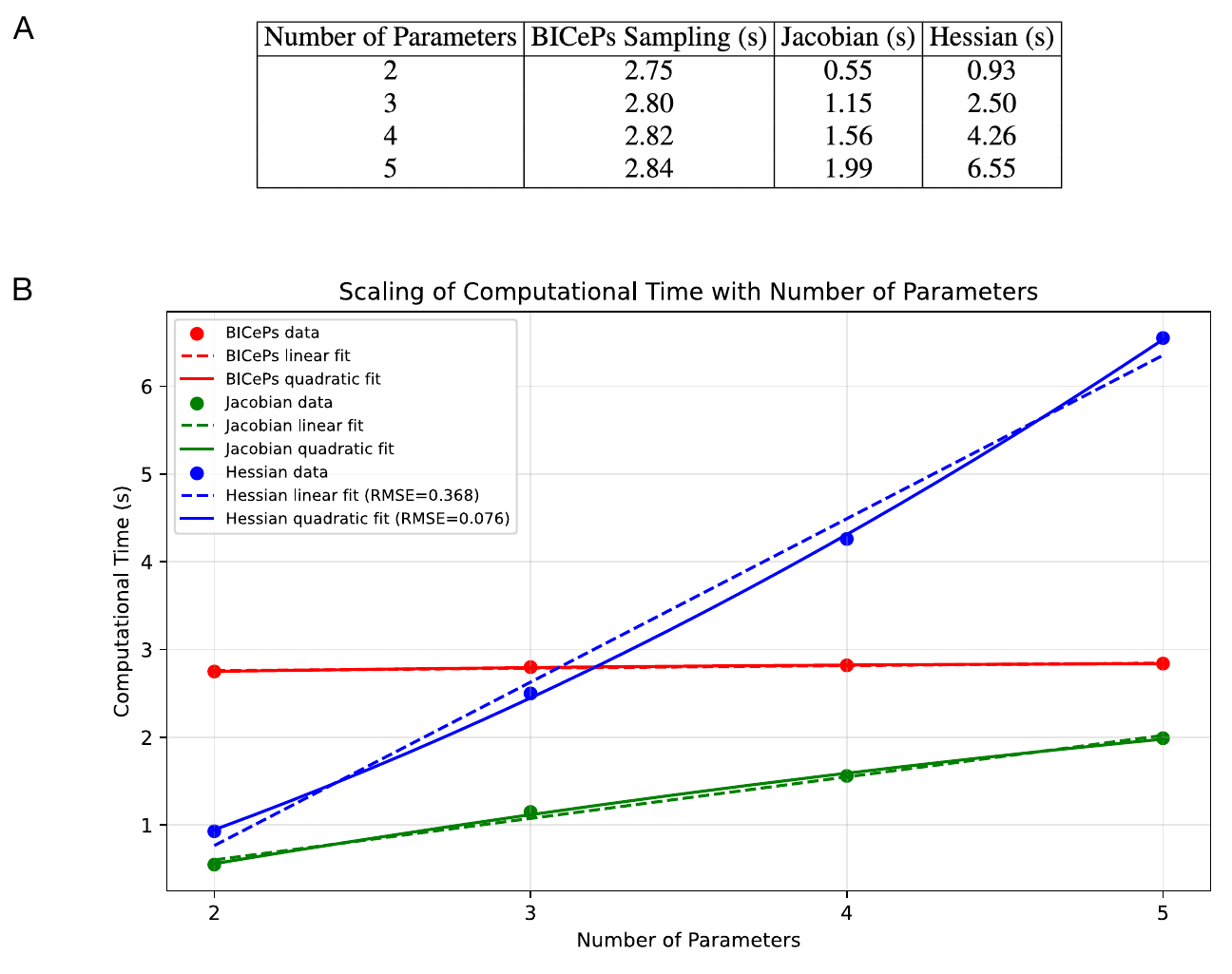}
  \caption{\small (A) Estimated computational runtimes for models with 2, 3,  4 and 5 parameters. All estimated runtimes were evaluated using a MacBook M1 Pro. (B) Linear and quadratic fits to the scaling data.  For the Hessian scaling data, a quadratic curve is a visibly better fit (RMSE = 0.368) compared to a linear fit (RMSE = 0.076).}
  \label{fig:runtime_eval}
\end{figure*}

\newpage